\documentclass[tradiabstract]{aa}

\usepackage{txfonts}
\usepackage{graphicx}
\usepackage{float}
\usepackage{natbib}
\usepackage[english]{babel}
\usepackage[hidelinks,colorlinks=true,linkcolor=blue,citecolor=blue]{hyperref}
\graphicspath{{/Users/kvoggel/Dropbox/Anil_Karina_Shared/Papers/NG7727/img/}}

\begin{document}

\title{First direct dynamical detection of a dual supermassive black hole system at sub-kiloparsec separation}

\author{Karina. T. Voggel\inst{1} \and Anil C. Seth\inst{2} \and Holger Baumgardt\inst{3} \and Bernd Husemann\inst{4} \and Nadine Neumayer\inst{4} \and Michael Hilker\inst{5} \and Renuka Pechetti\inst{6} \and Steffen Mieske\inst{7} \and Antoine Dumont\inst{2} \and Iskren Georgiev\inst{4}}

\institute{
Universite de Strasbourg, CNRS, Observatoire astronomique de Strasbourg, UMR 7550, 67000 Strasbourg, France, \email{karina.voggel@astro.unistra.fr}
\and University of Utah, James Fletcher Building, 115 1400 E, Salt Lake City, UT 84112, USA
\and School of Mathematics and Physics, University of Queensland, St. Lucia, QLD 4068, Australia
\and Max-Planck-Institut f\"ur Astronomie, K\"onigstuhl 17, 69117, Heidelberg, Germany
\and ESO Headquarters, Karl-Schwarzschild-Str. 2, 85748 Garching bei M\"unchen Germany
\and Astrophysics Research Institute, Liverpool John Moores University, 146 Brownlow Hill, Liverpool L3 5RF, UK
\and ESO Vitacura, Alonso de C\'ordova 3107, Vitacura, Casilla 19001, Santiago de Chile, Chile}
\date{Received 18. March 2021 / Accepted 31. October 2021}

\abstract{We investigated whether the two recently discovered nuclei in NGC\,7727 both host a super-massive black hole (SMBH). 
We used the high spatial resolution mode of the integral-field spectrograph MUSE on the VLT in adaptive optics mode to resolve the stellar kinematics within the sphere of influence of both putative black holes. We combined the kinematic data with an HST-based mass model and used Jeans models to measure their SMBH mass. 
We report the discovery of a dual SMBH system in NGC\,7727. We detect a SMBH in the photometric center of the galaxy in Nucleus 1, with a mass of $M_{\rm SMBH}=1.54^{+0.18}_{-0.15} \times10^{8}M_{\odot}$. In the second nucleus, which is 500\,pc offset from the main nucleus, we also find a clear signal for a SMBH with a mass of $M_{\rm BH}=6.33^{+3.32}_{-1.40}\times10^{6}M_{\odot}$. Both SMBHs are detected at high significance. The off-axis nature of Nucleus 2 makes modeling the system challenging; however, a number of robustness tests suggest that a black hole is required to explain the observed kinematics. The SMBH in the offset Nucleus 2 makes up 3.0\% of its total mass, which means its SMBH is over-massive compared to the $M_{\rm BH}-M_{\rm Bulge}$ scaling relation. This confirms it as the surviving nuclear star cluster of a galaxy that has merged with NGC\,7727. This discovery is the first dynamically confirmed dual SMBH system with a projected separation of less than a kiloparsec and the nearest dynamically confirmed dual SMBH at a distance of 27.4\,Mpc. 
The second Nucleus is in an advanced state of inspiral, and it will eventually result in a 1:24 mass ratio SMBH merger. Optical emission lines suggest Nucleus 2 is a Seyfert galaxy, making it a low-luminosity Active Galactic Nuclei (AGN). There are likely many more quiescent SMBHs as well as dual SMBH pairs in the local Universe that have been missed by surveys that focus on bright accretion signatures. }


\titlerunning{First Direct Dynamical Detection of a Dual Super-Massive Black Hole System at sub-kpc Separation}

\authorrunning{Voggel et al.}

\maketitle

\section{Introduction} \label{sec:intro}

When two galaxies with SMBHs in their centers merge, they are expected to form a dual SMBH with a separation of $1-100$ kpc (e.g., see \citealt{DaRosa2019} for a review). Dynamical friction and the ejection of stars
hardens the SMBH system until it reaches pc-scale separation and eventually merges via the emission of gravitational waves   \citep{Enoki2004, Makino2004, OLeary2009}. The low-frequency gravitational waves of such merging SMBH binaries will be the prime targets for upcoming space-based gravitational wave detectors such as LISA \citep{Moore2015}. The growth of SMBHs via galaxy mergers is a crucial ingredient in the assembly of BH masses in hierarchical galaxy evolution models \citep{Volonteri2003, Volonteri2010, Greene2012}. 

Active Galactic Nuclei (AGN) signatures have been used to identify a handful of dual SMBH candidates, with an even lower number being confirmed \citep{Komossa2003, Comerford2013, Comerford2015, Rubinur2017}. These candidates are typically at large distances (z>0.2), which make spatially separating their emissions challenging. Recently, large catalogs of candidate dual AGNs have been published \citep[e.g.]{Stemo2020}, but all of these dual AGNs have separations of 2-20\,kpc and are very distant. 
To confirm dual AGNs, several studies have used IFU spectroscopy to spatially resolve the emission line regions of each one \citep[e.g.,][]{Fu2012, McGurk2015, Kosec2017,Husemann2018}. However, at these distances, the most advanced integral-field-unit (IFU) instruments can identify the presence of two independent AGN components, but not the stellar component surrounding the SMBHs in the sphere of influence (SOI).

There is only one candidate for a dual AGN at sub-kpc separation with a dual Quasar pair and a projected separation of only 800\,pc \citep{Woo2014}.  The identification of sub-kpc SMBH pairs is especially important as it is the phase just before the formation of a gravitationally bound SMBH binary and its subsequent coalescence into a SMBH merger.

The closest known dual SMBH is NGC6240 \citep{Komossa2003, Kollatschny2020}, which is located at 144\,Mpc. 
In the study of NGC\,6240 by \cite{Kollatschny2020}, based on MUSE observations, a spatial resolution was reached that is good enough to show that the emission lines come from two separate stellar components and are thus a true dual SMBH. They discovered that it might be a triple AGN system with three accreting black holes within a separation of $\sim$1\,kpc. This galaxy is too distant for a dynamical BH mass measurement, and they can only infer the SMBH masses from the $M-\sigma$ relation. These are particularly uncertain given the dynamical state of the system. 
 
In this paper, we present a study of NGC\,7727, a candidate dual SMBH system. It lies at a distance of 27\,Mpc, and its disturbed post-merger morphology indicates it experienced a recent merger. The galaxy has a second nucleus with a candidate SMBH that was discovered by \cite{Schweizer2018}. It lies at a projected distance of just 500\,pc (3.8") from the main nucleus and has the same radial velocity as the main galaxy and is thus not a foreground or background object. Their integrated velocity dispersion measurement was higher than expected from a typical stellar population alone, indicating that there might be a SMBH in the second offset nucleus, which would make this a stripped nucleus in formation. 

The mass ratio of dual SMBHs is a key parameter for understanding GW signatures. Thus far, BH masses in dual SMBHs have been estimated using AGN indicators \citep{Greene2007, Schulze2015} or using the M-$\sigma$ relation \citep{Greene2007b, Xiao2011}. These estimates are uncertain, particularly due to the nonequilibrium conditions stripped galaxies show during merging (e.g., \citealt{Forbes2014}). Unfortunately, the large distances of most dual SMBHs make direct dynamical measurements of the SMBH masses impossible. Thus, NGC7727, as the nearest dual SMBH, represents a unique opportunity to make dynamical mass measurements and directly constrain the mass ratio of a dual SMBH system for the first time.

The AGN signatures detected by \cite{Schweizer2018} suggest Nucleus 2 may be a stripped nuclei. Dynamical detection would confirm this. Recently, stripped nuclei with SMBHs in their centers have been confirmed multiple times in massive star clusters in the halo of larger galaxies \citep{Seth2014, Ahn2017, Ahn2018, Afanasiev2018}. Their SMBHs were all confirmed through direct stellar dynamical measurements that resolved the sphere of influence of the SMBHs. These over-massive SMBHs make up between 3 and 15\% of the total mass, much more than what is found with the $M_{\rm BH}-M_{\rm Bulge}$ relation \citep[e.g.,][]{Kormendy2013}.  However, all these stripped nuclear star clusters are far out in the halo of their parent galaxies, and their former host galaxies were dissolved a few Gyr ago. 
In contrast, NGC\,7727 still shows active post-merger signatures and has been caught in the process of merging, making it an important test bed for SMBH formation as well as stripped nuclei formation. 

Simulations have predicted that the merging of galaxies produces stripped galaxies whose SMBHs are over-massive for the bulge mass of their host galaxy \citep{Volonteri2008, Volonteri2016,Barber2016,Tremmel2018}. The observed over-massive SMBHs in stripped nuclei are a confirmation of this theoretical expectation.

This paper is organized as follows. In Section \ref{sec:data}, we present the MUSE data and the analysis of the kinematics. In Section \ref{sec:three}, we present the photometric data and the mass modeling.
In Section \ref{dynmodel}, we present the results from the dynamical Jeans models that we used to measure the BH masses. In Section \ref{sec:AGN}, we discuss the findings concerning the emission lines in the galaxy. Finally, Sections \ref{sec:discussion} and \ref{sec:conclusion} contain a discussion and conclusion.

\section{MUSE data and kinematic measurements} \label{sec:data}
The center of NGC\,7727 was observed with the integral-field spectrograph MUSE \citep{Bacon2010} on the nights of the 8 and 29 July 2019 (Program ID: 0103.B-0526(A), PI: Voggel). For this we used the $7.5\arcsec\times7.5\arcsec$ narrow-field mode of the instrument that provides full adaptive optics and a pixel size of $0.025\arcsec$/pixel. The adaptive optics provides a $<0.1\arcsec$ point-spread function (PSF, 13\,pc at this distance), which is necessary to resolve the stellar velocities inside the $0.8-0.15\arcsec$ SOI of the two putative black holes.

\begin{figure}
   \centering
   \includegraphics[width=\hsize]{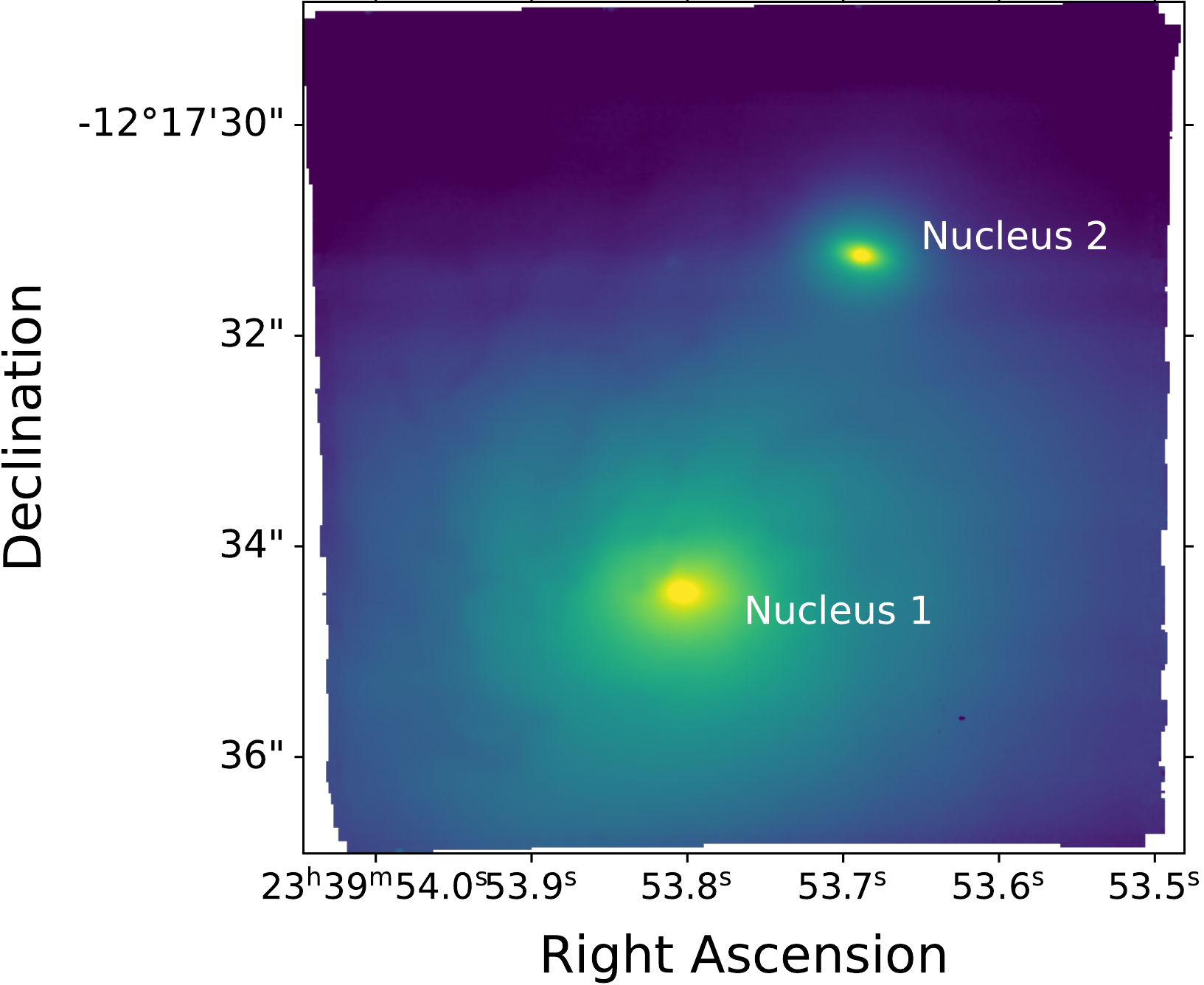}
     \caption{White-light image of the collapsed MUSE data cube. Nucleus 1 is the photometric center of the main galaxy NGC\,7727, and Nucleus 2 is offset to the north-west, which is 500\,pc in projected separation. } 
         \label{fig:FOV}
   \end{figure}

We observed two OBs with $4\times600$\,s observing time each, for a total time on target of 80 minutes. The individual cubes were fully reduced and then combined with the MUSE pipeline version 2.8.1 \citep{Weilbacher2020}. The image of the collapsed white-light cube is shown in Figure \ref{fig:FOV} with the two nuclei labeled. Nucleus 1 is at the photometric center of NGC\,7727, whereas Nucleus 2 is offset by $3.8\arcsec$ to the top right.
We did not use the provided sky subtraction of the pipeline due to the galaxy covering most of the field of view, which generally results in poor subtraction.  We chose the Zurich Atmosphere Purge (ZAP, \citealt{Soto2016}) package instead, which is a sky subtraction tool for MUSE based on a principal component analysis. This package has shown to preserve the line shape and flux of the lines while removing the sky efficiently, especially in cases where the objects fill the field of view of a cube.

\begin{figure}
   \centering
   \includegraphics[width=\hsize]{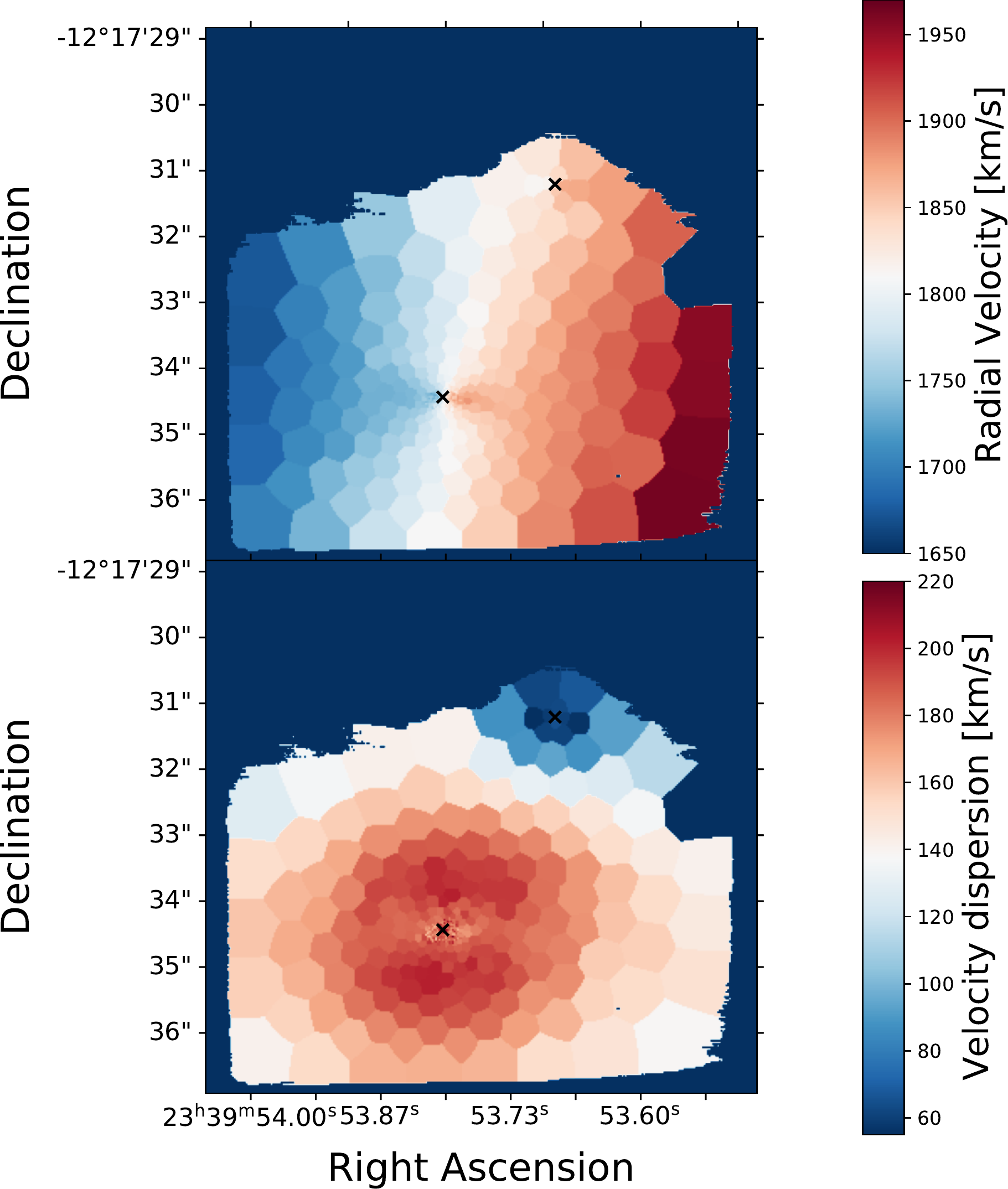}
     \caption{Top Panel: Stellar velocities are shown across the binned MUSE cube. Regions without sufficient S/N were excluded from the fits. Bottom Panel: Stellar dispersion. Both panels have the data in spatial bins of at least $S/N=25$, and the two Nuclei are marked with a black x. } 
         \label{fig:kin}
   \end{figure}

Due to MUSE's small NFM pixels ($0.025\arcsec\times0.025\arcsec$), we needed to spatially bin the spectra in areas where the S/N is too low. We did this using the Voronoi binning code \citep{Cappellari2003}, which creates bins with a similar minimum S/N. We binned to a S/N ratio of 20 for the region around Nucleus 2 to retain individual pixel resolution close to its center. To describe the kinematics in the center of Nucleus 1, a binning with a target S/N of 25 is sufficient for retaining the spatial resolution needed for dynamical modeling. For each of the Voronoi bins (see Fig. \ref{fig:kin}), we then combine the spectra within each bin into a single one-dimensional spectrum that then can be further analyzed.
We fit the binned 1-d spectra using the penalized pixel fitting code (pPXF) \citep{Cappellari2004, Cappellari2017}. We used the MUSE spectral library as stellar templates \citep{Ivanov2019}. The set contains 35 standard stars observed with MUSE cover stellar temperatures from 2600 and 33000\,K, $log(g)=0.6-4.5$ and  [Fe/H] from -1.22 to 0.55. Choosing a MUSE template library minimizes the template mismatch that can occur when using theoretical libraries that need to be convolved with the line-spread functions. 

In addition to these stellar templates, we also allowed for a gas emission component in the fits as the two galaxy nuclei show emission lines in the work of \cite{Schweizer2018}. The gas emission lines that are included in the fits are H$_{\rm \beta}$, H$_{\rm \alpha}$, [SII]$\lambda6716$, [SII]$\lambda6731$, [OIII]$\lambda5007$, [OI]$\lambda6300,$ and [NII]$\lambda6583$. We derived a flux for each emission line, and a single common gas velocity and dispersion that is independent of the stellar component. 

For the kinematic measurements, we restricted the MUSE wavelength range from 6500-9000\AA \,\,in order to closely match the wavelength of our mass model that is in the same wavelength range. We derived the uncertainty of the pPXF kinematic measurements using 50 Monte Carlo resamplings of each spectrum, adding a Gaussian error, and then refitting the kinematics. The standard deviation of the 50 fit results is used as 1-$\sigma$ error for the kinematics. The dispersion and velocity map of the entire field of view at the S/N$=25$ binning are shown in Fig. \ref{fig:kin}. The large-scale kinematic signatures of the galaxy are easily visible with strong rotation along the axis aligned closely north to south. While at this level of binning the details of Nucleus 2 are not resolved, it is clearly visible that the nucleus is at the same radial velocity as the surrounding stellar velocities of the main galaxy and thus definitely part of NGC\,7727.

\section{Photometric data and mass modeling}\label{sec:three}

\subsection{HST imaging and light profile }\label{sec:Galfit}
We used the F814W and F555W band observations of NGC\,7727 from the Hubble Space Telescope (HST) archive. We ran the light profile fitting code GALFIT \citep{Peng2004} on the F814W image to measure the light profiles and their structural parameters. GALFIT also requires the PSF of the observations, which we derived using the Tiny Tim software \citep{Krist2011}.

We fit the light profiles using GALFIT. The main galaxy was fit with three S\'ersic components: the bulge, Nucleus 1, and a S\'ersic component at the position of Nucleus 2. We excluded all regions with a significant contribution of dust from the fit. The results of the GALFIT run are listed in Table \ref{tab:galfit}. All magnitudes are expressed in the AB magnitude system with a zero point of $ZP_{\rm F814W}=24.21$.

\begin{table*}
\caption{\label{tab:galfit}GALFIT best-fit parameters}
\centering
\begin{tabular}{cccccc}
\hline\hline
Component & Mag$_{\rm 814}$  & $r_{\rm eff}$[\arcsec]  & S\'ersic Ind.  & Axis Ratio & P. A. \\
\hline
Bulge  & 11.28  &  6.2  &  2.28 &   0.74 &  -73.03 \\
Nucleus 1   & 15.39   &   0.36  &  1.51  &  0.60  &  82.57 \\
Nucleus 2  &  16.08  &    0.23  &  1.79 &   0.62  &  70.51 \\
\end{tabular}
\end{table*}
These light profiles were then de-projected into Gaussians with a multi-Gaussian expansion (MGE) \citep{Cappellari2002}. The MGE for Nucleus 1 de-projects the bulge component of NGC\,7727 as well as the Nucleus 1 profile itself. The de-projection of Nucleus 2 only contains the MGE of that component. The tables listing the MGEs are listed in the appendix. The standard de-projection of the MGE method only allows for components with the same center, and thus we do not include a component for the light of the main galaxy. We chose to directly use the light profiles for the Gaussian decomposition instead of the HST image itself because it provided us with the properties of the individual physical components of the galaxy. It also enabled us to fit for the main galaxy and Nucleus 2 at the same time, which is not possible with the MGE code. The typical relative residuals between the HST image and the light  profile are within 5\%. The MGE decomposition of this light profile has maximum relative errors of 9\% when decomposing the profile into individual Gaussian components.

\subsection{Determining the adaptive optics PSF}\label{sec:kinpsf}
The point-spread-function (PSF) of the MUSE adaptive optics (AO) correction needs to be measured precisely in order for us to model the kinematic signature of the potential black holes. In previous works \citep[e.g.,][]{Ahn2017, Ahn2018, Voggel2018}, we used HST imaging that has a higher spatial sampling than typical IFU instruments and convolved the HST images with a PSF model until the convolved light profile matched the one from the adaptive optics data. However, the MUSE narrow-field mode has a $0.025\arcsec$ pixel scale, which is twice the resolution compared to the $0.05\arcsec$ pixel scale of the available NGC\,7727 HST observations. Thus, we need an alternative route to determine the adaptive optics PSF. 

The light profile fit described in the previous section provide an analytic light profile for the Nuclei and the main galaxy, and thus we can use these light profiles to provide information on the shape beyond the resolution limit of the underlying HST data. We used the derived light profile parameters and created an un-convolved mock observation of NGC\,7727. We did this by combining a two-dimensional S\'ersic profile for each light profile component into a mock cube that has the MUSE spatial resolution. 

\begin{figure}
   \centering
   \includegraphics[width=\hsize]{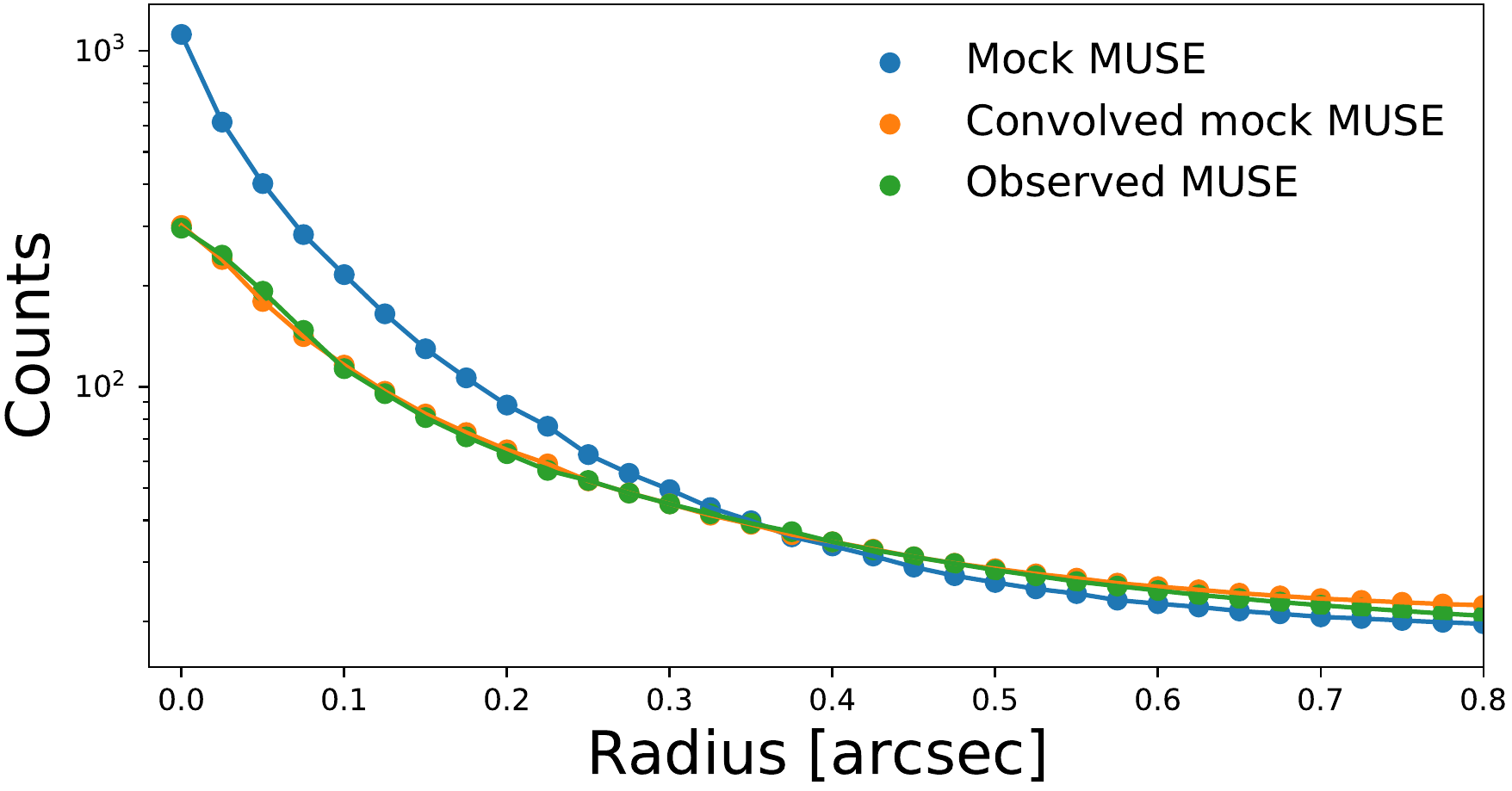}
      \caption{Radial brightness profile of Nucleus 2 is shown for the un-convolved mock cube in blue, the mock cube convolved with the best-fit AO PSF at 7500\AA\  is shown in orange, and the actual observed luminosity profile with MUSE is shown in green.} 
         \label{fig:radial_psf}
   \end{figure}

We then convolved these mock cubes with a wide grid of MUSE PSFs derived from the Maoppy code of \citet{Fetick2019}. In their model, they created a parameterization of the adaptive optics PSFs for MUSE using a Moffat core and Kolmogorov halo for the outskirts. The best-fit PSF minimizes the $\chi^{2}$ between the convolved mock image and the real MUSE observation light profile as shown in Fig. \ref{fig:radial_psf}.

   \begin{table}
\caption{ \label{tab:psf} Gaussian parameters for the kinematic PSFs. The first four components are for the best-fit PSF at 6000\,\AA, the second set is for the default 7500\,\AA \, PSF, and the last set is for the 8000\,\AA\ PSF.}
\centering
\begin{tabular}{ccc}
\hline\hline
Number & $\sigma$ ["]  & Fraction of the light [\%] \\
\hline
1 & 0.020 & 18.90 \\
2 & 0.099 & 4.83 \\
3 & 0.275 & 27.06 \\
4 & 1.03 & 48.65 \\
\hline
1 & 0.020 &  25.30 \\
2 & 0.137 & 2.78 \\
3 & 0.212 & 20.93  \\
4 & 0.891 & 40.78 \\
\hline
1 & 0.022 & 37.10 \\
2 & 0.098 & 15.23 \\
3 & 0.198 & 6.89 \\
4 & 0.758 & 40.07 \\
\end{tabular}
\end{table}

We derived our default AO PSF at 7500\AA, \,which is the mean wavelength of the MUSE spectra from which we derive the kinematics.  We chose the best-fit PSF by minimizing the difference between the convolved mock cube and the MUSE data. This best-fit PSF has a fried parameter of $r_{0}=0.2$, and $\alpha=0.9$, $\beta=1.9,$ which parameterize the Moffat core. The JAM models require a Gaussian parameterization of the input adaptive optics PSF, as the code cannot use the actual PSF. Thus, we expanded the parametrized PSF into 4 Gaussian components that represent the light profile of the PSF as close as possible. Four Gaussian components were necessary to provide a good fit to the model PSF. 
The sigma and relative light contributions of each Gaussian component are given in the first section of Table \ref{tab:psf}. The default 7500\AA \, PSF has two core components with a combined full width at half maximum (FWHM) of $0.07\arcsec$ that contains $\sim28\%$ of the light, the medium-sized component 3 contains about 21\% of the light, and the large $1\arcsec$ FWHM component contains $\sim$41\% of the light. This very concentrated core with a wide halo is typical for MUSE NFM PSFs for which core FWHMs of $0.07-0.1\arcsec$ are common. 

\begin{figure}
   \centering
   \includegraphics[width=\hsize]{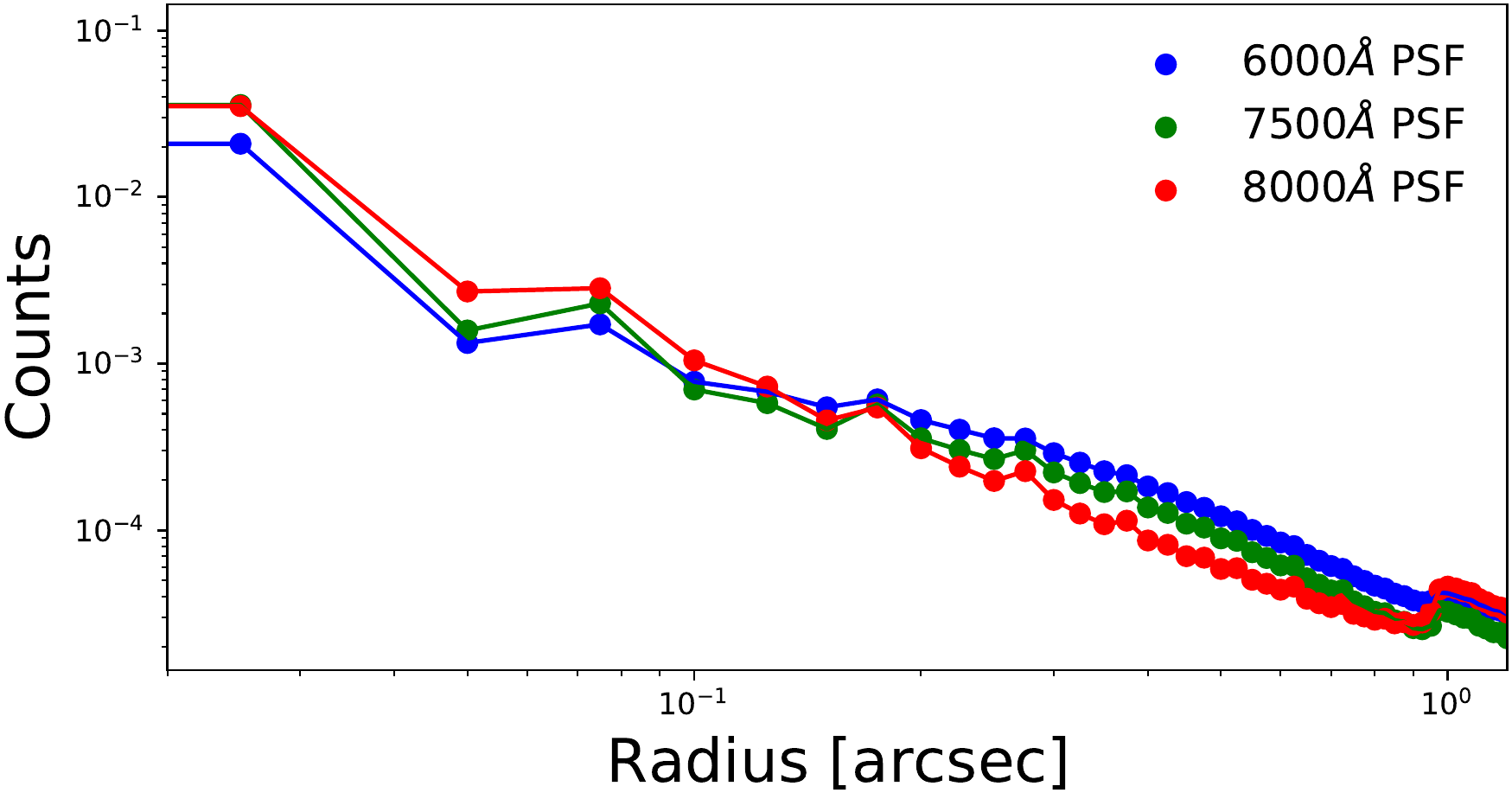}
      \caption{Comparison of the three MUSE AO PSFs we derived using the maoppy code. This is a radial profile taken from the full 2D PSF that the maoppy code provides on the same 0.025$\arcsec$ pixel size as the MUSE data. The variation of the data points are due to the MUSE pixel size in which the PSF was generated.} 
         \label{fig:psf_comparison}
   \end{figure}

The quality of the AO correction is wavelength dependent, with the best correction attained at red wavelengths. For the default PSF we assumed that the PSF follows the mean wavelength of the spectra at 7500\AA. However, to test for the systematic effects this assumption has on the SMBH masses, we also derived the AO PSF at 6000\AA\, and 8000\AA\,. In Section \ref{sec:systematics},
we test our models with those two PSFs and in Table \ref{tab:psf} their Gaussian expansion values are listed. A radial profile comparison of those two PSFs to the default 7500$\AA$ PSF is shown in Fig. \ref{fig:psf_comparison}. From the plot it is visible how the fraction of light in the red PSF at 8000\AA\, is more centrally concentrated, and for the blue PSFs their wings become much more extended.       

\subsection{Stellar population model}\label{sec:population}

\begin{figure}
   \centering
   \includegraphics[width=\hsize]{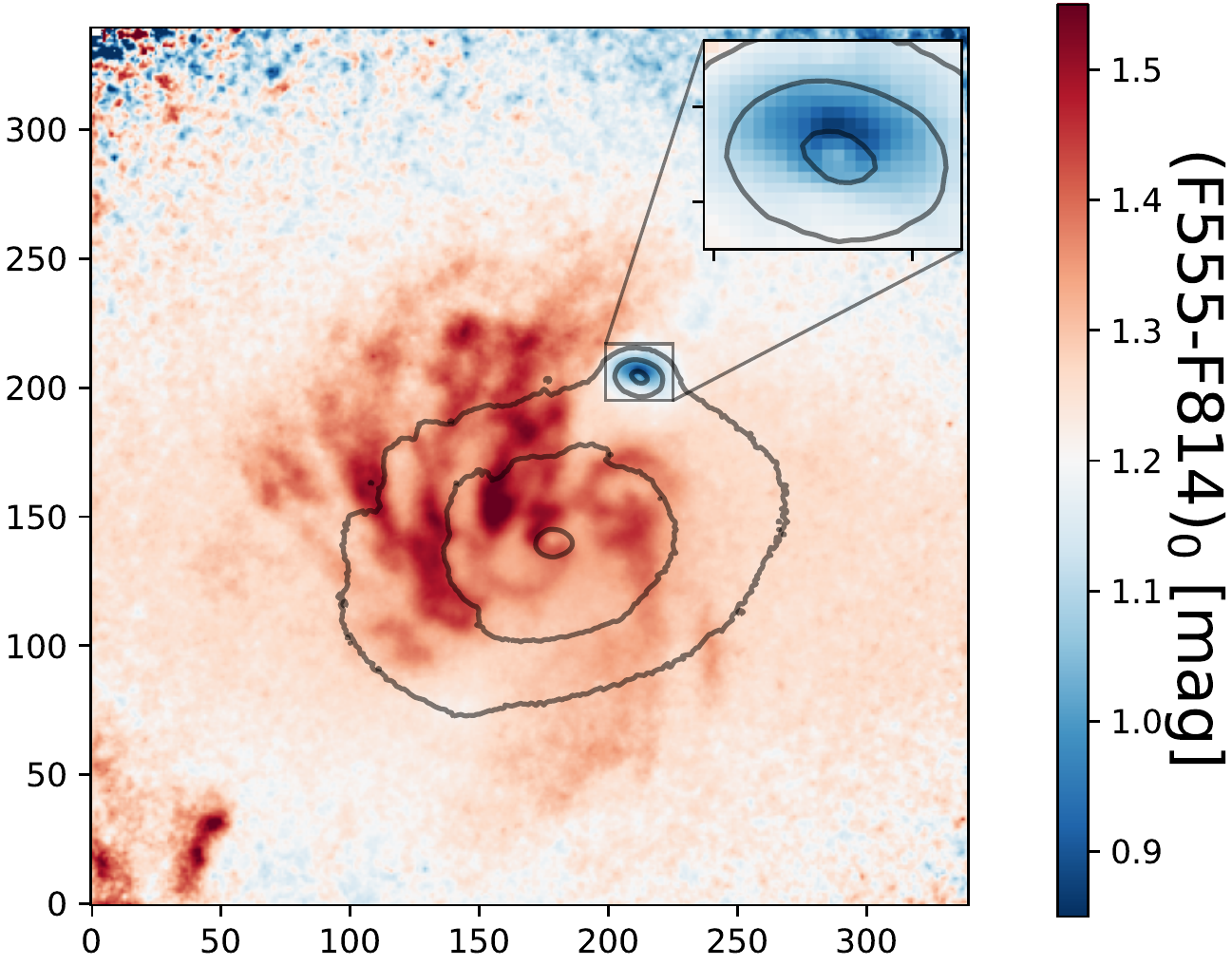}
      \caption{Map of F555W-F814W color derived from the HST images that are cross-convolved with the PSF of the opposite band. The color is corrected for foreground extinction. Inset focuses on Nucleus 2. The overlaid contours are from the F814 image.} 
         \label{fig:hst_colormap}
   \end{figure}

In \cite{Schweizer2018} it was found that Nucleus 2 is a mix of an intermediate stellar population with an age of 1.4\,Gyr and an older population of 11.3\,Gyr. We used the HST imaging to investigate whether these intermediate populations cause an age and color gradient that needs to be taken into account when modeling their stellar population. We used the archival HST imaging in the F555W and F814W bands to create a color map. For this color map, we cross-convolved each image with the PSF of the opposite band to avoid any color effects due to PSF differences. We then corrected for foreground extinction. The resulting color map is shown in Fig.\ref{fig:hst_colormap} with overlaid intensity contours of the F814W HST image. It is easily visible that Nucleus 2 is much bluer than the entire surrounding galaxy and Nucleus 1. The prominent dust lanes also point to some significant amount of dust and thus internal extinction in the galaxy.

The center of Nucleus 2 has a color of F555W-F814W=1.0\,mag and then rises outwards. This indicates a strong central concentration of a different population than the main galaxy. From the zoomed-in image in the inset of Fig.\,\ref{fig:hst_colormap}, it is also visible that the color north of the center of Nucleus 2 is much bluer than south of it. This asymmetric color is likely caused by dust. Interestingly, in the very central two pixels of Nucleus 2 the color becomes redder again, indicating there is some dust contribution in the very center.

To take this color gradient into account for our mass models that are required as input for the Jeans models, we used single stellar population (SSP) fits to determine whether there is also a mass-to-light ratio (M/Ls) gradient. For the stellar population fits we used the MILES stellar population library \citep{Vazdekis2010,Vazdekis2011} with Basti isochrones and a Chabrier IMF. We only used non-alpha-enhanced models in our fits, for which in the MILES models, [Fe/H]=[M/H]. The fits were performed on radially integrated spectra of the MUSE data with a step width of $0.05\arcsec$ that corresponds to two MUSE pixels. This allowed us to have better S/Ns for the population fits. To minimize the impact of the wavelength varying PSF, we restricted the population fits to include only the blue wavelength range of $4800-5770$\,\AA, where there are several age sensitive lines. 

Other than restricting the wavelength range and using a theoretical library of stellar templates, the setup of the rest of the fitting routine is the same as in the section of the dynamical measurements. The spectral variation in the PSF prevents the derivation of extinction from the spectroscopic data. We used the weights that the pPXF code provides for each stellar template to calculate the weighted average of the radial $M/L_{\rm I}$ and the predicted $F555W-F814W$ color without accounting for internal extinction.

The observed foreground extinction-corrected radial HST color profiles of Nucleus 1 and 2 are compared to the predicted color from our SSP fits (diamond symbols) in the top and middle panels of Fig. \ref{fig:SSP_radial}, respectively. For Nucleus 2, the HST color profile is derived from the upper half of the nucleus in order to obtain a better estimate of the intrinsic profile that is less affected by the dust in the southern region. 

\subsection{Internal extinction correction}
In both cases, the observed HST colors are significantly redder by $0.1-0.2$ mag compared to the predicted colors. This reddening is likely caused by internal dust and thus extinction in NGC\,7727, which has to be taken into account for our mass modeling. We can use the difference between the observed HST color to estimate the internal extinction in both nuclei and how it varies with radii. For Nucleus 1, the color difference does not significantly depend on radius, suggesting a roughly constant internal extinction. However, in Nucleus 2 the observed color is significantly reddened in the center then drops around its half-light radius and then rises again. We calculated the internal extinction by first transforming the F555W-F814W color excess into E(B-V), and then using the \cite{Calzetti2000} extinction law with a $R_{\rm V}=3.1$ to calculate the internal extinction. We transfer to the HST bands using the relation $A_{\rm F814W}/A_{\rm V}=0.605$. 

The effects of the HST PSF are known to cause an artificial drop of the colors to the blue in the 1-2 most central pixels (e.g., \citealt{Ahn2018}). As this is the crucial area for our SMBH determination, we need to carefully calculate this PSF effect on our observed HST color profile.
To assess by how much the center of our profile is bluer, we use the unconvolved mock cube from the earlier section and convolve it with the F814W PSF and once with the F555W PSF to mimic an object with constant color profile. The resulting radial F555W-F814W profile drops by 0.06\,mag toward the blue in the central $0.05\arcsec,$ even though we assumed a constant color. Beyond that radius, the color profile is not significantly affected. To correct this PSF artefact, we added a 0.06\,mag reddening to the central value of the observed HST color profile. We then proceeded to derive the overall extinction profile from the derived reddening, with this PSF effect already corrected in the observed HST profile.

\subsection{Mass-to-light profiles}
\begin{figure}
   \centering
   \includegraphics[width=\hsize]{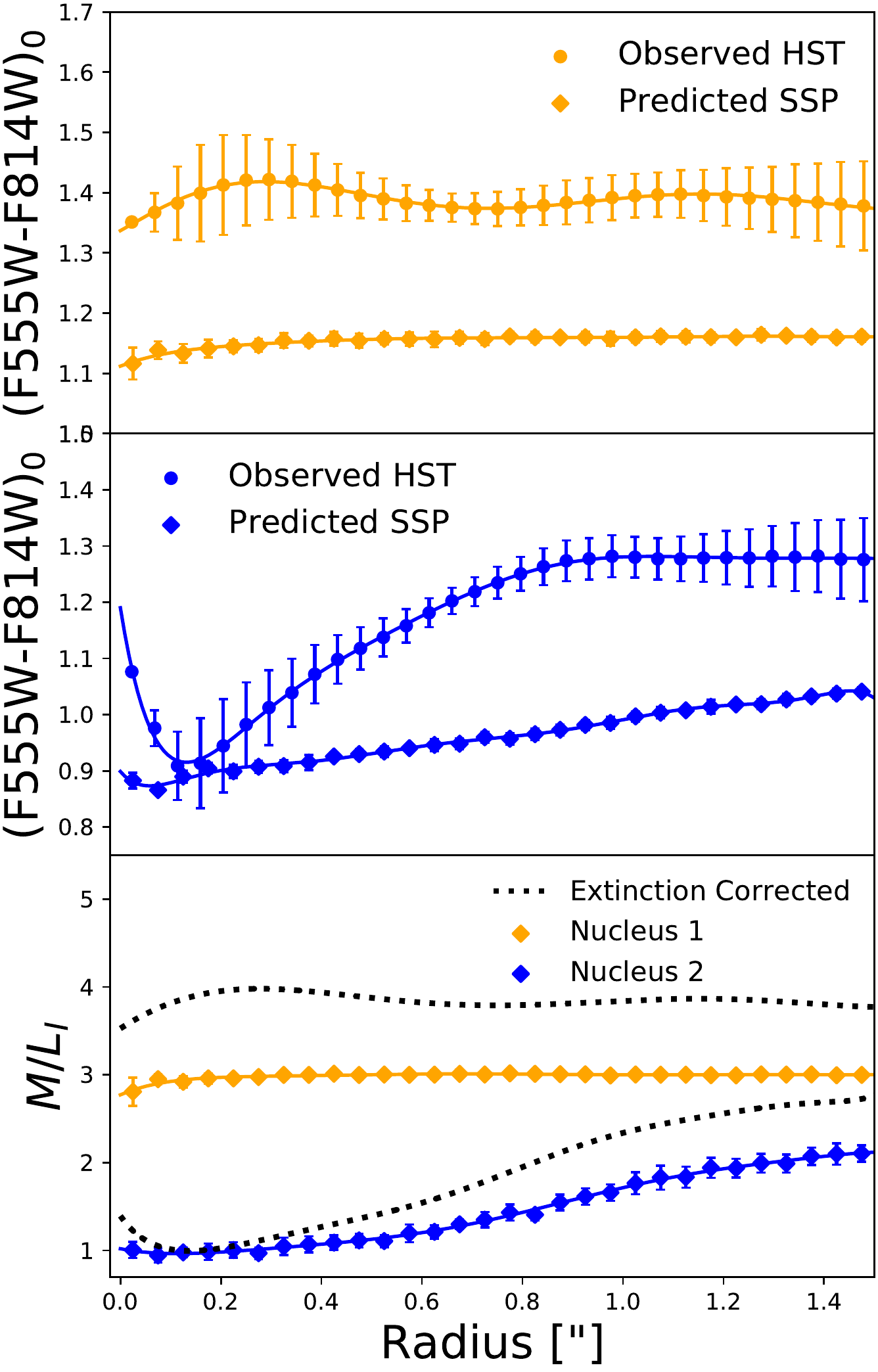}
      \caption{Radial observed HST $F555W-F814$ color profiles are compared to the predicted color profiles in the top panel for Nucleus 1, and in the middle panel for Nucleus 2. The observed HST colors that were corrected for foreground extinction and PSF effects. The relative reddening of the observed profiles is used to calculate the radially dependent internal extinction of the two Nuclei. \textbf{Bottom Panel:} Colored diamond symbols are the predicted radial $M/L_{\rm I}$ profiles for both nuclei, and the dashed black lines are the $M/L_{\rm I, eff}$ profiles that have been corrected for internal extinction. } 
         \label{fig:SSP_radial}
   \end{figure}

In the bottom plot of Fig. \ref{fig:SSP_radial} the predicted $M/L_{\rm I}$ profiles from the SSP fits are shown. The effective $M/L_{\rm eff}$ that is corrected for internal extinction is shown via black dashed lines. We correct the radial M/L profiles with the internal $A_{\rm F814W}$ extinction that was calculated in the previous section based on the colour difference between the observed HST colors and the predicted model colors. The effective M/L is then calculated via $M/L_{\rm eff}=M/L_{\rm I}*10^{0.4*A_{\rm F814W}}$.
 
The SSP fits clearly show that there is a M/L gradient in Nucleus 2. This is caused by our stellar fits finding that Nucleus 2 is dominated by  a young and centrally concentrated stellar population. In the center of Nucleus 2, the composition of stellar templates mostly consists of two young 0.8 and 1.5\,Gyr and metal rich ([Fe/H]=0.06-0.4) templates with a smaller contribution of an intermediate-age 6\,Gyr template. On average, the young population has an age of $3.25\pm0.5$\,Gyr, whereas the old stellar population has an average age of $10.24\pm1.8$\,Gyr. This confirms that the overall color gradient to redder colors at larger radii is due to an age gradient, not a metallicity difference. No metal poor templates below $[Fe/H]=0.06$ are used in any of the fits. Only at larger distances above $0.8\arcsec$ does the contribution of the old stellar population of NGC\,7727 become significant.

The significant gradient in $M/L_{\rm I,eff}$ on Nucleus 2 necessitates the incorporation of a variable M/L into our mass models. 
The standard JAM models just allow for one constant M/L ratio otherwise. Variable M/L profiles have been used in several previous works that measured the black hole masses in nuclei \citep[e.g.,][]{Ahn2018, Nguyen2017, Nguyen2018, Nguyen2019}.
We added the population gradient to the JAM models by multiplying our MGE mass models with the extinction-corrected $M/L_{\rm I}$ profile. The $M/L_{\rm I}$ that corresponds to a given MGE Gaussian component is calculated at the radius of the standard deviation of that MGE component. In the JAM models, instead of fitting for a M/L ratio, we fit a mass-scaling factor $\Gamma$ that scales the radial mass profile according to the pre-determined stellar mass gradient.

In comparison, Nucleus 1 has no such strong M/L gradient, and its M/L and color profile are near constant within the central $1.5\arcsec$. The templates' stellar fits to the spectra show a consistently old $12.45\pm0.18$\,Gyr and metal-rich population with [Fe/H]=0.4\,dex.

\section{Dynamical modeling}\label{dynmodel}
To model the black hole masses, we used Jeans anisotropic models (JAMs) \citep{Cappellari2008, Cappellari2020}. These models solve the stellar Jeans equations to predict the velocity and dispersion fields based on a mass model of the galaxy. The JAM models need the AO PSF and the stellar-mass profile of the galaxy components that we derived in Section \ref{sec:three} as inputs. The JAM models solve the Jeans equations to predict the second-order velocity moment $v_{\rm rms}=\sqrt{v^{2}+\sigma^{2}}$, and they then integrate the same spatial bins we used in our MUSE data. 

A cut-out of the observed $v_{\rm rms}=\sqrt{v^{2}+\sigma^{2}}$ map in the very centers of Nucleus 1 and 2 are shown in the leftmost panel of Fig. \ref{fig:obsmodel_nuc1} and Fig. \ref{fig:obsmodel_nuc2}, respectively. These $v_{\rm rms}$ maps are the input that is given to the dynamical models. 

Our Jeans models have the BH mass, mass-scaling factor $\Gamma$, anisotropy $\beta,$ and inclination as free parameters. Usually, the mass-scaling factor is equal to the M/L ratio, but as we include the $M/L_{\rm eff}$ gradient that was determined in Section \ref{sec:population} this factor does not represent a single M/L value, but rather a scaling of the M/L gradient. We used a Markov chain Monte Carlo (MCMC) sampler to efficiently explore this parameter space. We used the emcee package \cite{emcee2013} which provides a python version of a MCMC sampler. We ran the MCMC chains with 20 walkers and 10 000 steps and discarded the first 150 steps of each walker as the burn-in phase.

\subsection{Nucleus 1} \label{sec:nuc1}
\begin{figure}
 \centering
 \includegraphics[width=\hsize]{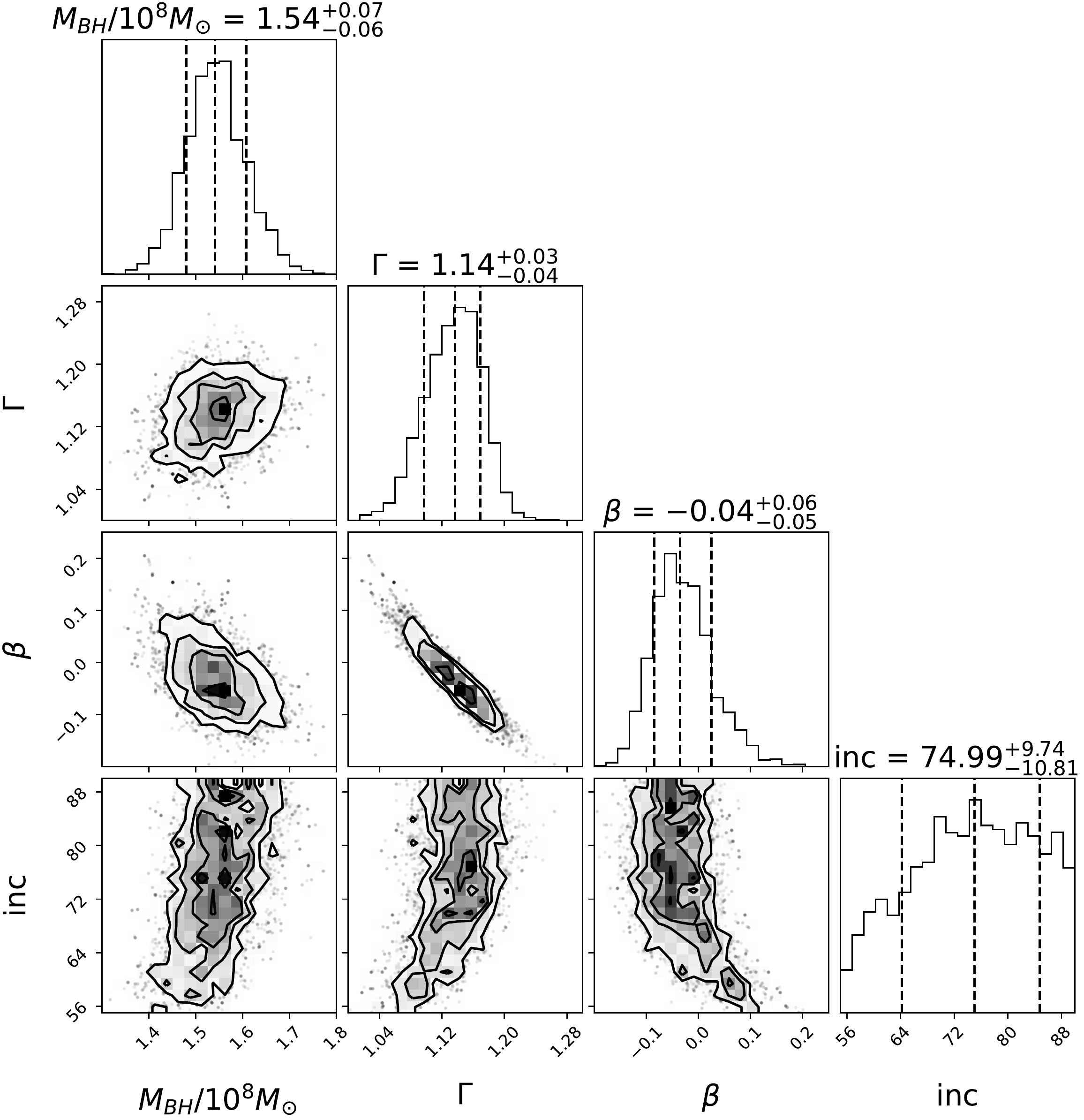}
  \caption{Corner plot shows the output of JAM models for Nucleus 1, from 10.000 MCMC steps, with the first 150 steps discarded as a burn-in phase. The best-fit BH mass, the mass-scaling factor $\Gamma$, anisotropy $\beta,$ and inclination angle are shown as marginalized histograms as well as their dependence on each other.} 
         \label{fig:corner_nuc1}
   \end{figure}

The MCMC Jeans model derives a best-fit black-hole mass for Nucleus 1 of $M_{\rm BH}=1.54^{+0.07}_{-0.06}\times10^{8}M_{\odot}$. The corner plot of the distribution of the MCMC realizations is shown in Fig. \ref{fig:corner_nuc1}. 
The model finds a best-fit anisotropy that is mildly tangential but still fully consistent with zero. The best-fit mass-scaling factor that scales the derived M/L profile shown in Fig. \ref{fig:SSP_radial} is $\Gamma=1.14^{+0.03}_{-0.04}$ , indicating that the main galaxy is made up of an old stellar population that has a high intrinsic $M/L_{\rm I}\sim3.4$, slightly higher than our expected value of around 3 (yellow diamond symbols, bottom panel of Fig. \ref{fig:SSP_radial}). The total stellar mass of the bulge plus nucleus of NGC\,7727 in our dynamical models is $M=5.24 \times10^{10}M_{\odot}$. The radius within which the dynamical model was compared to the observed $v_{\rm rms}$ maps is $0.5\arcsec$, which contains 276 individual bins.

\begin{figure*}
 \centering
 \includegraphics[width=\hsize]{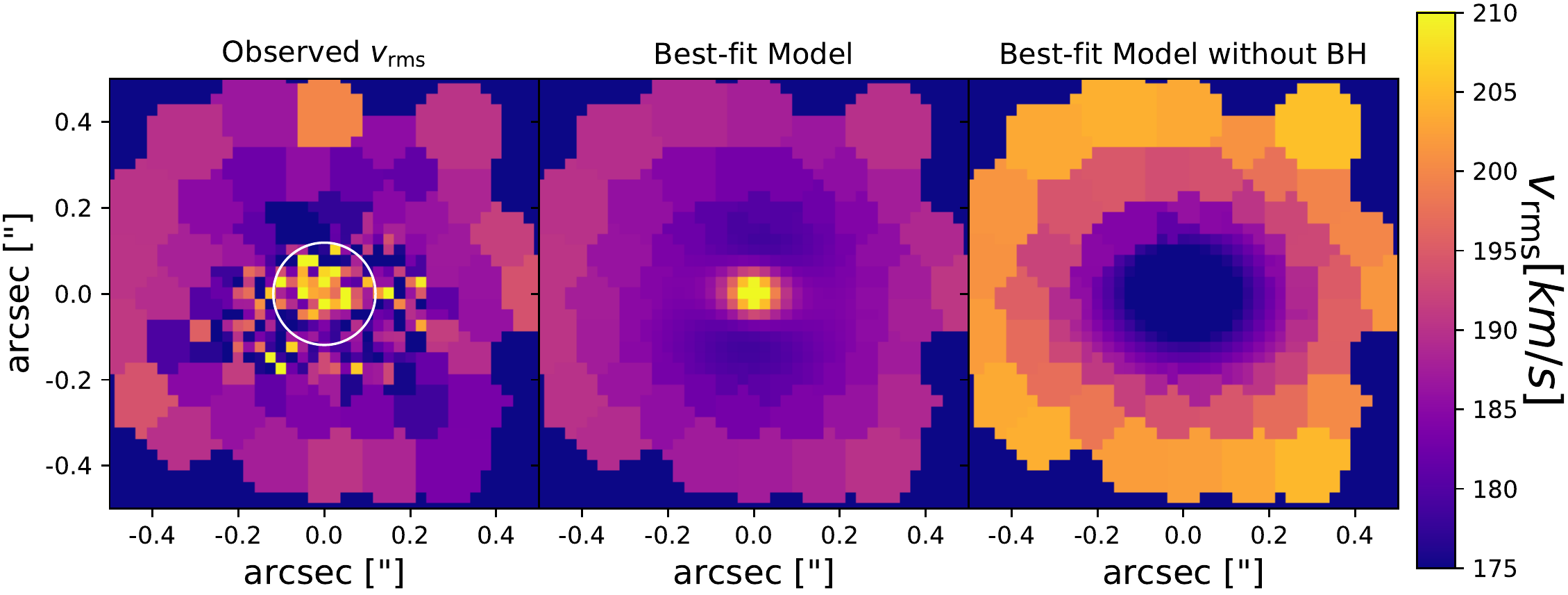}
  \caption{ $v_{\rm rms}$ maps centered on Nucleus 1. Left panel shows the observed MUSE map, including a circle that marks the sphere of influence. In the middle is the best-fit model, and on the right is the best-fit model for a MCMC run when not allowing for a black hole at all. The typical errors on the $v_{\rm rms}$ values in the MUSE map are 4\,km\,s$^{-1}$.} 
         \label{fig:obsmodel_nuc1}
   \end{figure*}

In Figure \ref{fig:obsmodel_nuc1}, the $v_{rms}$ map of the observed MUSE data is shown on the left, the best-fit model is in the middle, and the best-fit model that does not allow for a SMBH is shown on the right hand side. The stellar-mass model on the right shows that without a BH in the center, the $v_{\rm rms}$ of NGC\,7727 decreases sharply, because the much lower velocity dispersion from the nucleus itself is now dominating the center. Instead, a central increase in $v_{\rm rms}$ is seen, due to the presence of the SMBH.
The theoretical SOI of the best-fit SMBH is given by  $\frac{G*M_{\rm BH}}{\sigma^{2}}=16.86$\,pc, which is equal to $0.12\arcsec$ or $\sim$5 MUSE pixels at this distance. The core component of the MUSE PSF having a FWHM of $0.07\arcsec$ shows that we easily resolved the SOI of this black hole. 

The reduced $\chi^{2}$ of the best-fit model is 1.36. The difference with the best-fit model that does not allow for a SMBH at all is $\Delta\chi^{2}=1631,$ illustrating that the model requires a SMBH at extremely high statistical significance.
In Section \ref{sec:systematics}, we show that our measurement errors are systematics dominated, but all models explored retain a significant black hole mass.

\subsection{Nucleus 2} \label{sec:nuc2}
Nucleus 2 is embedded in NGC7727, which has a much higher velocity dispersion and thus can artificially raise the inferred velocity dispersion in Nucleus 2. Current state-of-the-art dynamical models cannot include off-center components, so we cannot model the main galaxy and Nucleus 2 at the same time. The best solution with these constraints is thus to evaluate the JAM models in a region where the contamination from the underlying galaxy's light is negligible.
To quantify this effect, we modeled the relative light contribution of the main galaxy at a given radius from Nucleus 2 and then modeled to what extent different relative galaxy contributions raise the velocity dispersion. 
 \begin{figure*}
 \centering
 \includegraphics[width=\hsize]{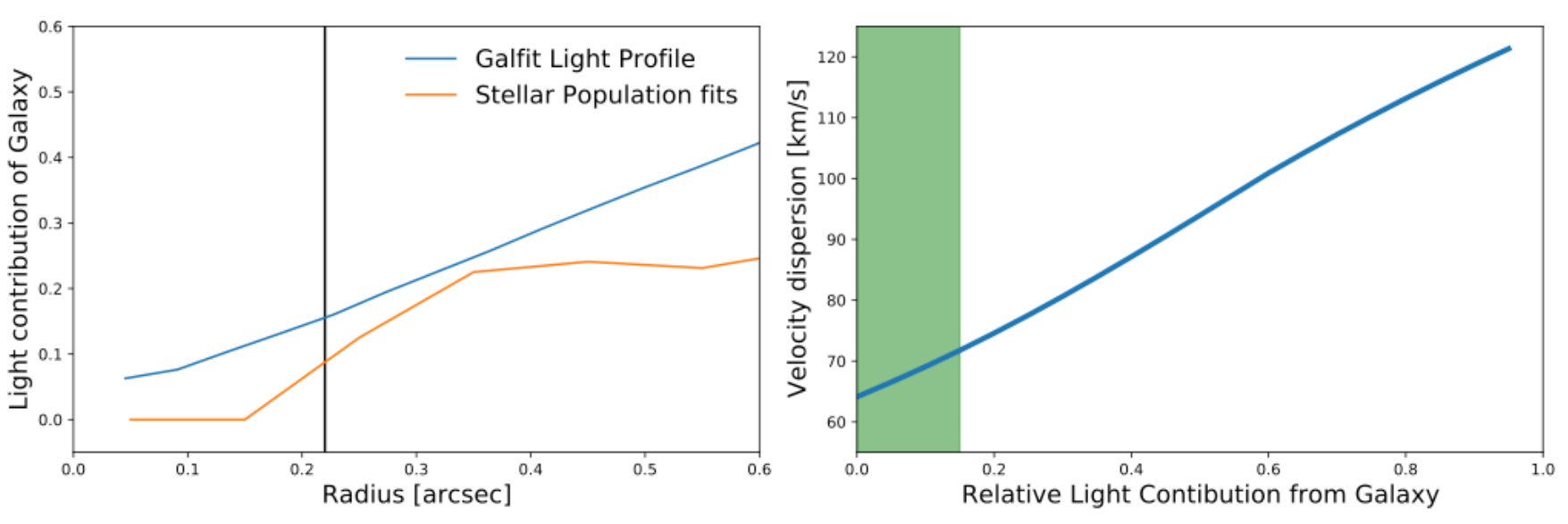}
  \caption{Relative light contribution of the main galaxy at the location of Nucleus 2. \textbf{Left:} Relative light contribution of the main galaxy shown as radial profile from the center of Nucleus 2: blue line is based on the individual light profiles, and once orange line is the relative contribution of old stellar population templates in the fits. The back vertical line marks the half-light radius of Nucleus 2. \textbf{Right:} Velocity dispersion derived with pPXF from spectra that contain increasing relative amounts of galaxy light. The green shaded region marks the relative contribution range we expect within the half light radius of Nucleus 2.} 
         \label{fig:disp_test}
   \end{figure*}  

Using the two-dimensional light profiles of each nucleus and the main galaxy from Section \ref{sec:Galfit}, we derived the relative light contribution of the main galaxy as a function of distance from Nucleus 2, which is shown in the left panel of Figure \ref{fig:disp_test}. This contamination of NGC7727 light decreases toward the center of Nucleus 2 and is <15\% within the half-light radius of Nucleus 2 (black vertical line). We did a similar test using the stellar population models from Section \ref{sec:population}, where we assumed that the light fraction assigned to templates with ages larger than 10\,Gyr come from the body of the galaxy. The main body of the galaxy is uniformly old with only templates above 12\,Gyr contributing. The stellar population fits (orange in Fig. \ref{fig:disp_test}) show that in the center of Nucleus 2 there is very little contribution of the main galaxy, and it is typically lower than the estimate from the light profiles. The two methods give a range of 5-10\% of typical galaxy contamination within Nucleus 2.

We now model how relative light contributions of the main galaxy increase the inferred velocity dispersion. We use a typical template spectrum for Nucleus 2 and convolve it with a Gaussian velocity dispersion of 65 \,km\,s$^{-1}$ and one representing the main galaxy convolved with the equivalent of 140\,km\,s$^{-1}$, the dispersion of the main galaxy at the distance of Nucleus 2. We then co-added these two spectra with relative light contribution of the main galaxy ranging from 0-95\% in 5\% increments. We ran pPXF on these artificial test spectra to determine their velocity dispersion. The results of that test are shown in the right panel of Fig. \ref{fig:disp_test}. We find that for galaxy light contaminations of up to 15\% (shaded area), the increase in velocity dispersion is below 5\,km\,s$^{-1}$. This is similar to our typical error bars on the $v_{rms}$ of 4-5\,km\,s$^{-1}$. Therefore, the effect of the main galaxy on the dispersion in the center of Nucleus 2 where we look for the BH signal should be small. In addition, the contamination increases the dispersion toward the outskirts of Nucleus 2, and it would thus decrease a typical SMBH signal rather than mimic a central rise. Therefore, this effect would only bias us toward a too low BH mass. We thus evaluate the JAM models within a radius $r<0.35\arcsec$, covering almost twice the half-light radius; moreover, they have 88 kinematic bins, of which the vast majority are within the half-light radius of Nucleus 2, where we expect very little galaxy contamination.

 \begin{figure}
 \centering
 \includegraphics[width=\hsize]{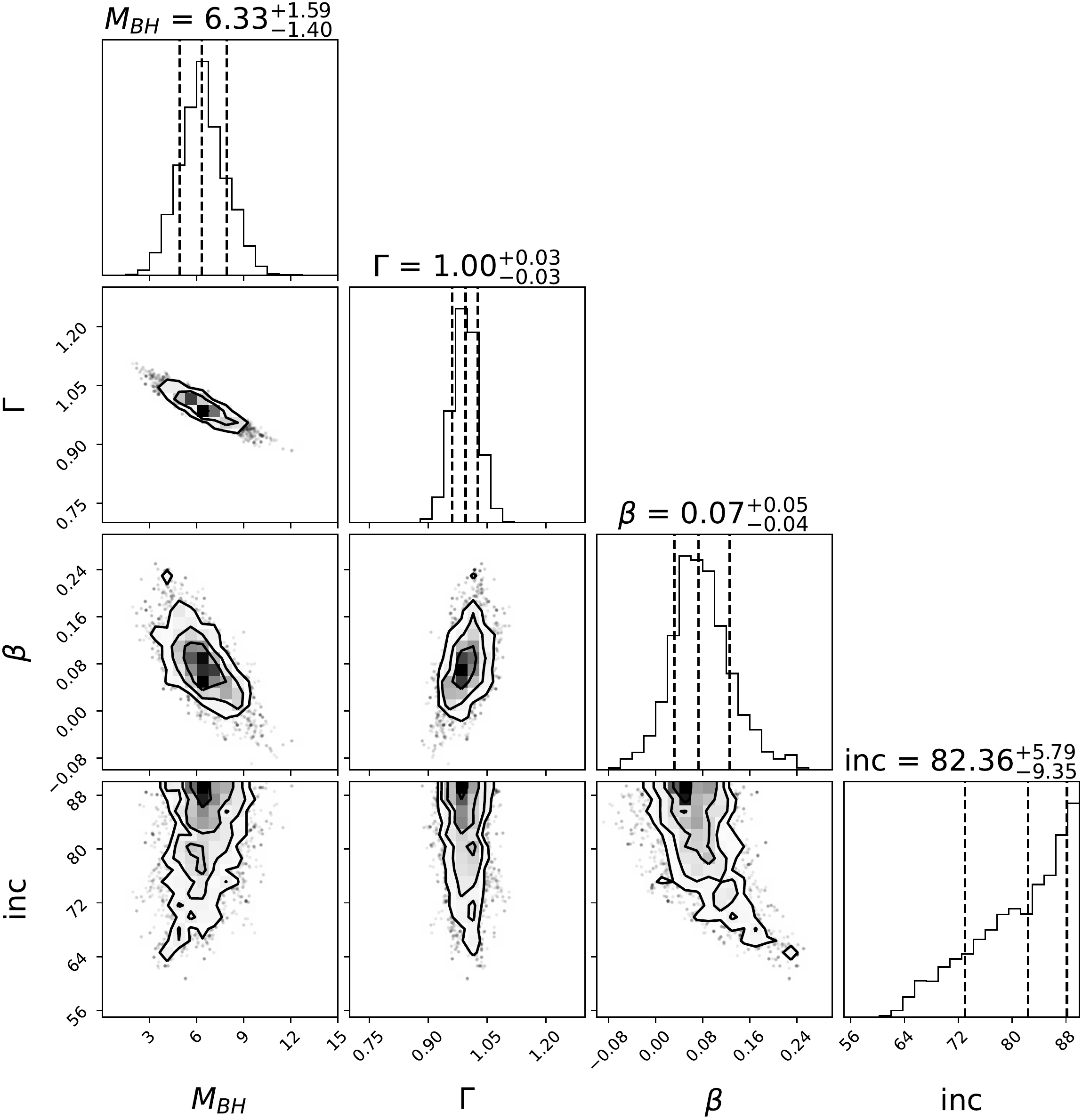}
  \caption{Corner plot showing the output of JAM models for Nucleus 2, from 10.000 MCMC chain steps with the first 200 steps of the burn-in discarded. The best-fit BH mass, the stellar mass scaling factor $\Gamma$, anisotropy $\beta,$ and the inclination angle are shown as marginalized histograms, and their dependence on each other is also shown.} 
         \label{fig:corner_nuc2}
   \end{figure}  
   
We derived a best-fit black hole mass for Nucleus 2 of $M_{\rm BH}=6.33^{+1.59}_{-1.40}\times10^{6}M_{\odot}$. The corner-plot with the results of the MCMC realizations is shown in Fig. \ref{fig:corner_nuc2}. The reduced $\chi^{2}$ of the best-fit model is 1.63, showing that we obtain a decent fit to our data. The statistical significance of the detection from the marginalized SMBH mass distribution is 4.5\,sigma. The best-fit $\beta=0.07^{+0.05}_{-0.04}$ shows that the nucleus exhibits a small level of anisotropy.
The anisotropy is degenerate with inclination, with lower values of anisotropy for edge-on inclinations.
The black-hole mass does not depend on the inclination, and only a small correlation with $\beta$ is observed. The best-fit mass scaling factor is $\Gamma=1.00\pm0.03,$ which is exactly what one would expect for an accurate mass model. 

\begin{figure*}
 \centering
 \includegraphics[width=\hsize]{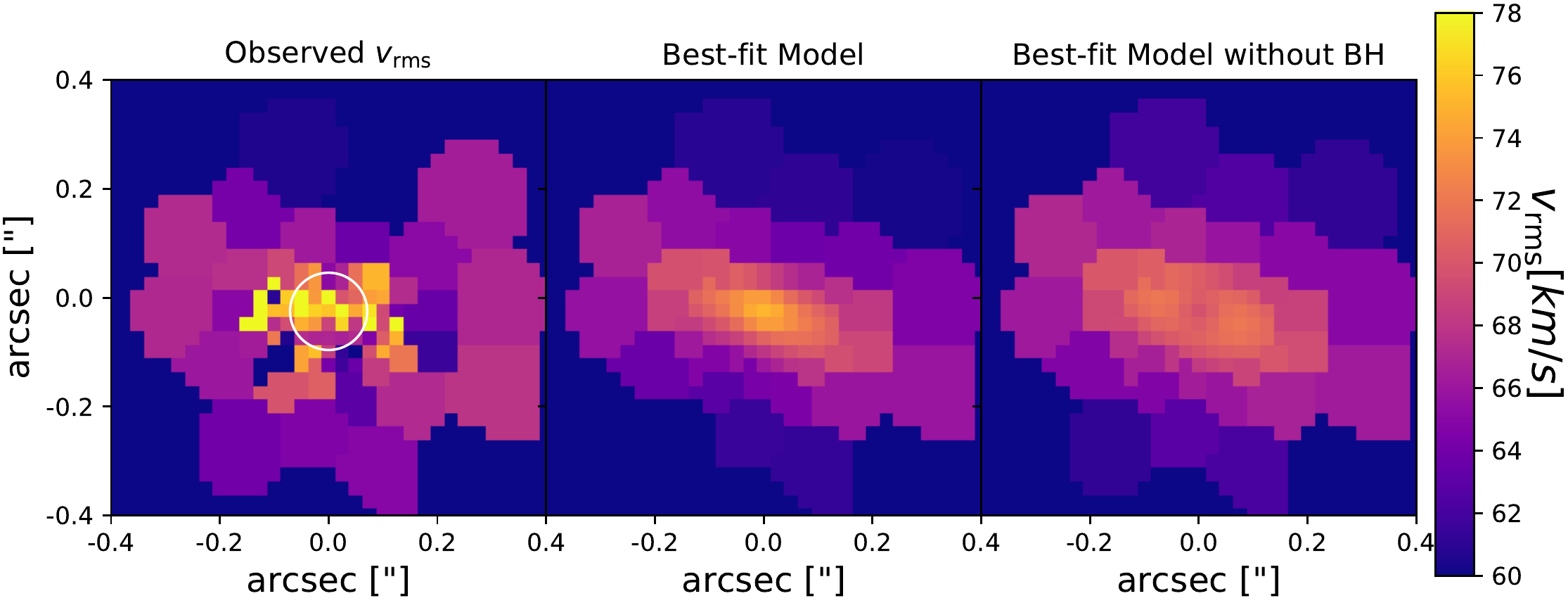}
  \caption{ $v_{\rm rms}$ maps centered on Nucleus 2. Left panel shows the observed MUSE map; in the middle is the best-fit model, and on the right is the same model but without a black hole. The white circle marks the SOI of $0.08\arcsec$. The typical errors on the $v_{\rm rms}$ values in the MUSE map are $4km\,s^{-1}$ , and thus while the observational scatter appears large it is consistent with the best-fit model in the center with a reduced $\chi^{2}$ of 1.6. We also note the small velocity scale of the colorer, which covers only   18\,km\,s$^{-1}$ to give sufficient contrast in displaying the models.}  
         \label{fig:obsmodel_nuc2}
   \end{figure*}

The statistical uncertainty on the MCMC run in Fig. \ref{fig:corner_nuc2} indicates that the BH mass is detected at $\sim5\sigma$ significance when assuming that our errors are Gaussian. However, to further test the significance, we ran a set of dynamical models where we forced a zero-mass SMBH. The best fit for the model without SMBH is worse by a $\Delta\chi^{2}=18.7$, ruling out the no black hole model at $8\times10^{-5}$ confidence level, indicating that a model containing a SMBH is strongly preferred. This way of assessing the statistical confidence is almost identical to the 4.5\,$\sigma$ level derived from the marginalized SMBH mass distribution.  The $v_{\rm rms}$ map of the observed MUSE data, the best-fit model, as well as the best-fit no SMBH model are shown in Figure \ref{fig:obsmodel_nuc2}. This comparison illustrates the need for a SMBH in the center of Nucleus 2. In the model not allowing for a SMBH, there is a decline in $v_{\rm rms}$ at the center, which is contrary to the observed rise. 
The SOI of this SMBH embedded in Nucleus 2 is SOI$=10.45$\,pc ($0.08\arcsec$ at this distance). This is within the resolution range of the two core components of the PSF component that have a combined FWHM of $0.07\arcsec$.

The total stellar mass of Nucleus 2 is $2.10\times10^{8}M_{\odot}$. The best-fit BH makes up 3.0\% of this stellar mass. 
Nucleus 2 has a stellar mass that is well below what was predicted in \cite{Schweizer2018}. This difference is likely due to the literature integrated dispersion being $\sigma=79\rm km\,s^{-1}$ and thus $12\rm km\,s^{-1}$ larger than the $\sigma=66.3 \pm 1.3\rm km\,s^{-1}$ we find as an integrated value for Nucleus 2. The slit used to take their spectra was $1\arcsec$ long and thus picked up a lot of galaxy light, which starts to dominate the light at $\sim0.7\arcsec$ from Nucleus 2, and our measurement integrates the light within the 0.3" circular aperture. This effect is visible in the right panel of Fig. \ref{fig:disp_test}, where we show how we tested the increase in velocity dispersion due to an increasing contribution of the main body of the galaxy. At a contribution of 50\% from the underlying galaxy, the predicted velocity dispersion would be $90km\,s^{-1}$ due to this contamination. For the small aperture of $0.3\arcsec,$ we predict a contamination of less than 10\%, which will raise the velocity dispersion by less than $5km\,s^{-1}$.

\subsection{Systematic uncertainties of the SMBH mass measurement}\label{sec:systematics}     
\begin{figure*}
 \centering
 \includegraphics[width=\hsize]{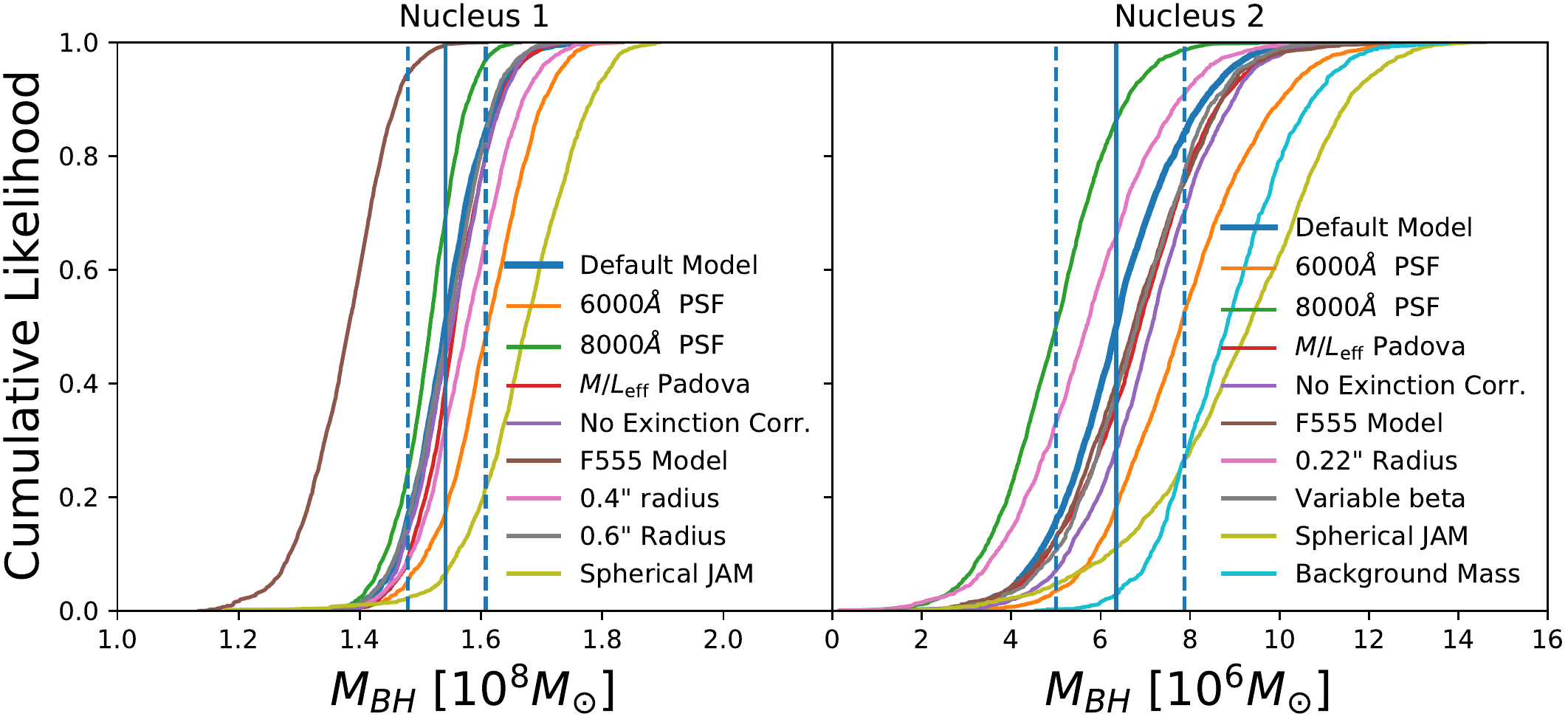}
  \caption{Cumulative likelihood distribution of BH masses for both black holes in Nuclei 1 and 2. The blue curve is the default model for both nuclei, and the dashed lines are the $1\,\sigma$ levels of the default model. The cumulative distribution of variations of the model inputs, such as the PSF and $M/L_{\rm eff}$ , are shown as colored distributions.} 
         \label{fig:cumu_like}
   \end{figure*}

In the last section we calculate the formal statistical errors of the BH mass measurement; however, models with a high number of data points can result in stronger constraints on the fit parameters, which makes it easier for the systematics to dominate. We investigated these systematic uncertainties, by varying the model assumptions we make in our analysis and quantifying the impact on the SMBH mass determination. The cumulative likelihood of all these tests is shown in Figure \ref{fig:cumu_like}.
In particular, we tested how the wavelength-dependent AO PSF influences our results by using PSFs derived at a wavelength of 6000\AA\, and at 8000\AA\, instead of the default PSF at 7500\AA. Their Gaussian expansion is shown in Table \ref{tab:psf}. In both nuclei using the bluer 6000\AA, the PSF increases the SMBH mass to just below the 1$\sigma$ level. 
When modeling the SMBH with the 8000\AA\, PSF, which is more centrally concentrated, the SMBH mass for Nucleus 1 decreases insignificantly, while for Nucleus 2 it decreases to around the $1\,\sigma$ limit. For Nucleus 2, the uncertainties derived from the PSF are clearly the dominating systematics and they exactly encompass the 1$\sigma$ range we derived with our MCMC runs.

To test the uncertainties in our mass models, we used the Padova \citep{Padova2000} instead of the Basti SSP models when deriving the input $M/L_{\rm I}$ profile. For both Nuclei, the change in stellar templates in our population determination has negligible effects on the SMBH mass and is consistent with the error bars.

We also tested how robust the JAM models are against our extinction correction. The runs without any extinction correction show little to no difference compared to the default models, and thus the models are not sensitive to this parameter. 

Another test is run by using the best-fit F555W Galfit models instead of the default model based on the F814W band. We chose the redder band in our default model, because these wavelengths are less affected by dust. 
For Nucleus 2, the change of band in which we model the stellar population has no effect on the SMBH mass determination. In Nucleus 1, however, the new model in F555W has a significantly smaller SMBH mass $(1.4\times10^{8}M_{\rm \odot)}$ than the default model. This change in SMBH mass is caused by the higher S\'ersic index of Nuclei 1 in the F555W model with n=1.99 versus 1.51 in the old F814 model. This uncertainty in its light profile is likely due to the significant dust contribution around Nuclei 1, which makes modeling it in F555W harder (and likely less accurate) than in F814W. We used this largest variation as lower boundary for our SMBH mass estimate.
This systematic uncertainty amounts to 9\% in the SMBH mass. Nucleus 1 has reached a regime where mass determination is dominated by systematics in the modeling and not by statistics of the observations, which are very small due to the high number of spaxels. 

Nucleus 2 shows an increase in its central M/L. We tested whether this could be due to radially varying anisotropy that could mimic the BH signal. For this, we assigned three different $\beta$ that were allowed to vary individually without being tied to each other. We assigned the four MGE components with radii below $0.05\arcsec$ to be $\beta_{1}$, the three MGEs from r$=0.1-0.3\arcsec$ $\beta_{2,}$ and the three outer component are characterized by $\beta_{3}$. The full MCMC run finds isotropy in the center with $\beta_{1}=0.01,$ and the two outer components are mildly anisotropic with $\beta_{2,3}=0.06$. The SMBH mass in this run is $6.82\times10^{6}M_{\rm \odot,}$  thus even higher than for a model that only allows for a single anisotropy. Therefore, we do not find any evidence for strong radial anisotropy that could mimic the signal of AMBH in the center of Nucleus 2, even when we allow it to vary radially. This finding is consistent with what has been found for Omega Cen in \cite{Zocchi2017}, where the center is isotropic and the intermediate parts are radially anisotropic.

We also tested whether our Nuclei are better modeled using JAM models, where the velocity ellipsoid is spherically aligned \citep{Cappellari2020}. In this case, the SMBH mass for Nucleus 2 is significantly higher in these spherical models, while the anisotropy becomes mildly tangential with $\beta$=-0.07. The tangential anisotropy leaves more room for a SMBH in the model, but as we typically do not observe nuclei or star clusters with high central tangential anisotropies, we did not adopt this model as our default. For Nuclei 1, using a spherical model similarly results in a significantly higher SMBH mass of $1.69\times10^{8}M_{\rm \odot,}$ while also leading to a significant tangential anisotropy of $\beta=-0.25$. Similarly to Nucleus 2, this high level of tangential anisotropy is not typically observed and is thus unrealistic.

Finally, we tested whether increasing or decreasing the fitting radius in which the JAM model is evaluated has any effect on the BH measurement in Nucleus 1. The default model uses a $0.5\arcsec$ limit, and we tested $0.4\arcsec$ and $0.6\arcsec$ as variations. For both fitting radii, the SMBH mass of Nucleus 1 is fully consistent with our default model.
Due to Nucleus 2 being off-center, we can only evaluate the region within 0.35\arcsec that is directly dominated by Nucleus 2 in the dynamical models, as explained in Section \ref{sec:nuc2}. Therefore, we only added a test with a smaller radius of 0.22\arcsec, which is equal to the half-light radius of Nucleus 2. The SMBH mass in that model is slightly smaller than our default but well within our $1\,\sigma$ uncertainties.

One limitation of JAM models is that they do not allow mass contributions that are off-center. To include the underlying mass of NGC7727 and its dark matter, we incorporated a MGE component in the surface potential of the Jeans models, which approximates their mass contribution as locally constant. This locally flat component can be achieved by adding a Gaussian MGE component with a larger radius than Nucleus 2 itself ($2\arcsec$ were used in this case). 

Using the surface density determined in the JAM models of Nucleus 1, we find the surface density of NGC7727 at the location of Nucleus 2 to be $1.4\times10^{4}M_{\odot}/pc^{2}$. As we have no large-scale kinematic data to measure the dark matter mass, we need to estimate its density at the location of Nucleus 2. NGC\,7727 has a total stellar mass of $M=1.3\times10^{11}M_{\odot}$ \citep{Schweizer2018}, and thus we expect it to reside in a $\sim10^{13}M_{\odot}$ dark matter halo assuming a typical stellar mass to halo mass relations \citep[e.g.,][]{Behroozi2013}. Using the stellar-to-halo radius relation \citep[e.g.][]{Jiang2019}, we estimate that the virial dark matter halo radius of NGC\,7727 should be $\sim$90\,kpc based on its stellar radius of $r_{\rm e}=3.7$\,kpc. With a concentration index of c=15 and the total halo mass, we calculate the predicted DM density assuming a NFW halo profile at 500pc from the galaxy center to be $2M_{\odot}/pc^{3}$. Projected, this is equal to a surface density of $80M_{\odot}/pc^{2}$, showing how the center of the galaxy is completely dominated by the stellar mass. We added these two surface densities to the gravitational potential in the JAM models of Nucleus 2. The results of this test with an additional background mass component (Fig. \ref{fig:cumu_like}) show that the best-fit SMBH is $8.3\times10^{6}M_{\odot}$ and thus larger than in our main models. Therefore, even when accounting for the galaxy mass in the background, the observed kinematics absolutely require a massive black hole. Without a SMBH a central drop in dispersion is seen and a SMBH is again required at high significance. 
The findings from our  $v_{\rm rms}$ test in Section \ref{sec:nuc2} showed the same effect: contamination by the background galaxy results in a modestly higher SMBH mass. Given that this effect we tested is the maximum effect (i.e., Nucleus 2 could be at larger physical radius than its projected radius) and offsets the BH mass by only slightly more than our 1\,$\sigma$ upper limit, we did not incorporate this component into our default model.

Besides the spherical JAM model and the background mass model, all other tested variations of the JAM model inputs of Nuclei 2 are consistent with the errors of the best-fit BH mass measurement, indicating that our statistical errors of 22\% are large enough to encompass typical uncertainties in our modeling. The dominant systematics are the knowledge of the AO PSF that encompass our 1$\sigma$ range. The spherical JAM model allows for a higher BH mass and is the only one that finds a tangential anisotropy, which we do not typically observe in other stripped nuclei \citep{Seth2014, Ahn2017, Voggel2018}. However, in the Milky Way and Cen\,A nuclear star cluster, there is some evidence for central tangential anisotropy near the SMBH \citep{Cappellari2009, Feldmeier-Krause2017}. The dominant systematics for the upper limits are the treatment of the background galaxy with our test showing up to a 25\% larger SMBH is possible. 
Thus, we quadratically add the largest systematic in the upper limit (spherical model) of Nucleus 2 quadratically to the error for a final SMBH mass of $M_{\rm BH}=6.33^{+3.32}_{-1.40}\times10^{6}M_{\odot}$.
It is important to note that none of the model variations allow for a zero BH model in any of the models, making our SMBH detection extremely robust against systematic effects.

For Nucleus 1, the F555 model and the spherical model lie outside the $1\,\sigma$ regime, indicating that we are in a regime that is dominated by systematic uncertainties in our modeling rather than statistics. The statistical errors are only 3.9\% of the SMBH mass in Nucleus 1. To provide more realistic errors and include systematic uncertainties in the SMBH mass of Nucleus 1, we added the statistical errors and the largest derived systematic uncertainty of err$_{\rm sys}=^{+0.17}_{-0.14}$ quadratically to Nucleus 1 for a final SMBH mass, with total uncertainties of $M_{\rm BH}=1.54^{+0.18}_{-0.15} \times10^{8}M_{\odot}$.

\section{AGN Signatures} \label{sec:AGN}

There is additional support for the presence of black holes in both nuclei of this galaxy from their emission lines. It was already reported in \citet{Schweizer2018} that Nucleus 1 is consistent with a low-ionization nuclear emission-line region (LINER) emission profile, but that work did not find significant X-ray emission at Nucleus 1 that pointed to an AGN origin. Nucleus 2 hosts an X-ray source with a luminosity of L=$2.8\times 10^{39}$\,erg\,s$^{-1}$ \citep{Brassington2007}.

\begin{figure}
   \centering
   \includegraphics[width=\hsize]{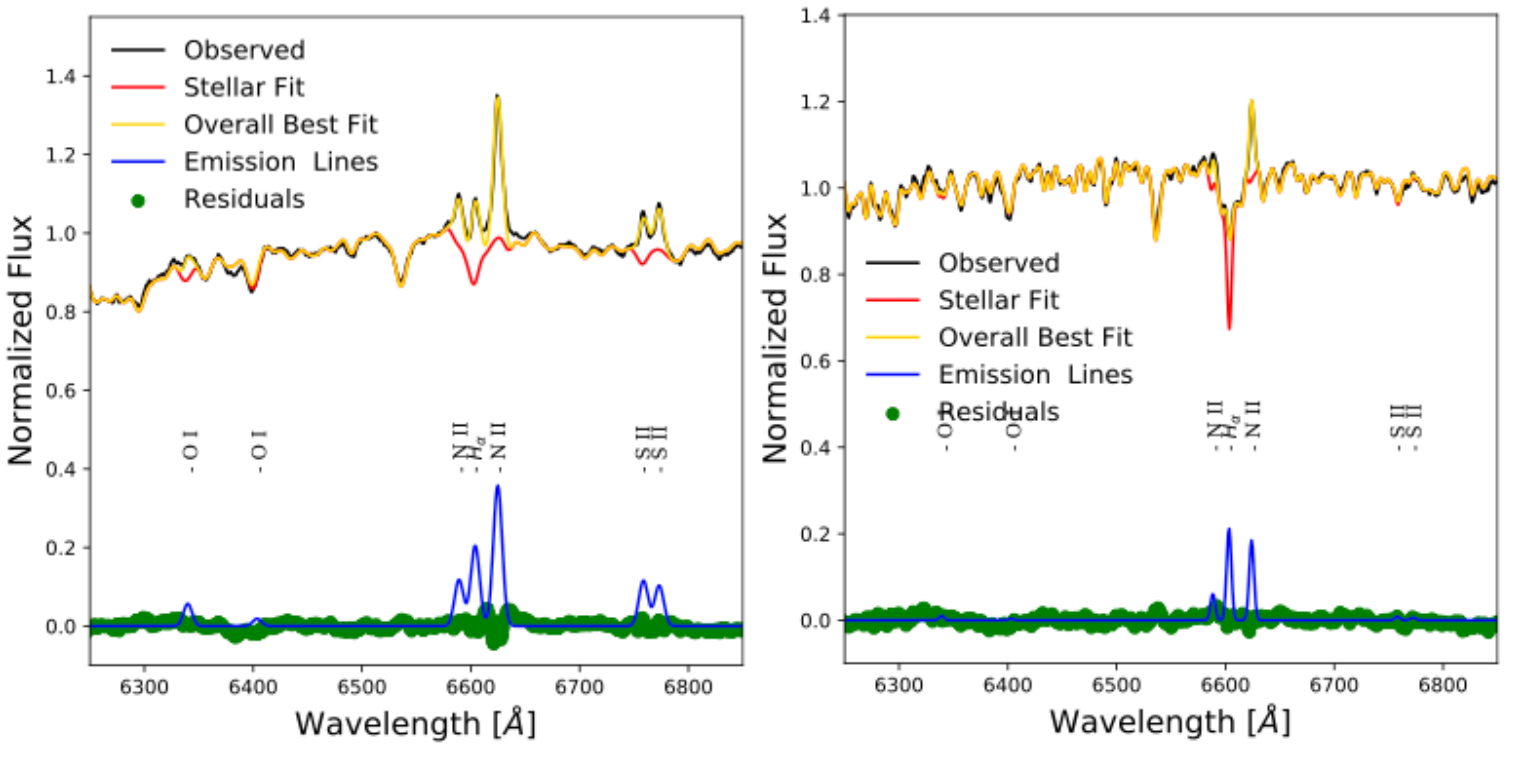}
      \caption{Part  of the integrated spectra of Nucleus 1 (left panel) and Nucleus 2 (right panel), showing a zoomed-in image of the area containing the [NII]$6583$, H$_{\alpha,}$ and both [SII] lines. The overall best pPXF fit is shown in yellow, the stellar-component is shown in red, and the emission lines are shown in blue.} 
         \label{fig:emission}
   \end{figure}

We investigated the emission-line ratios of both nuclei by plotting them on the diagnostic BPT diagram. This diagnostic plot is commonly used to discriminate between AGN and star-forming regions \citep{Kewley2001}. We used the line fluxes as determined by pPXF in Section \ref{sec:data}, where we describe simultaneously fitting the stellar and gas components. The code provides the fluxes of the individual lines within the following spectral range: H$_{\rm \beta}$, H$_{\rm \alpha}$, [SII]$\lambda6716$, [SII]$\lambda6731$, [OIII]$\lambda5007$, [OI]$\lambda6300,$ and [NII]$\lambda6583$. The region of the MUSE spectra that contains [NII]$\lambda6583$, H$_{\rm \alpha,}$ and both [SII] emission lines is shown in Figure \ref{fig:emission} for the integrated spectra (within $r<0.2\arcsec$) of Nuclei 1 and 2, respectively. 

\begin{figure*}[!ht]
   \centering
   \includegraphics[width=\hsize]{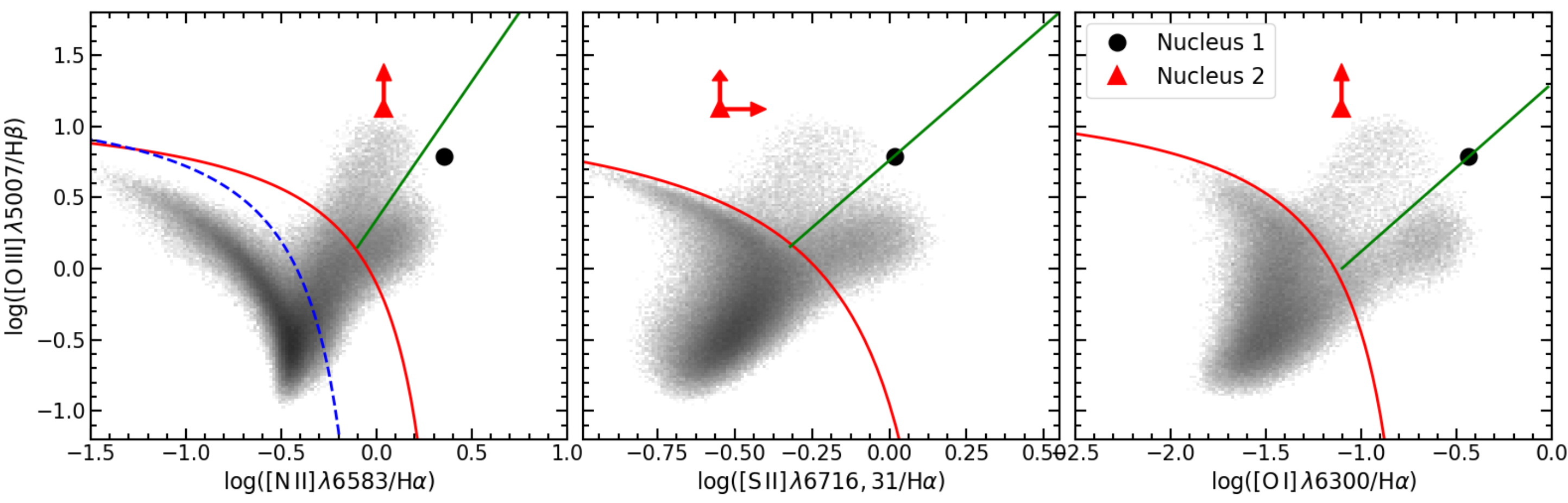}
      \caption{Three diagnostic BPT diagrams for Nucleus 1 (black point) and Nucleus 2 (red triangle).  The gray data points are taken from the  SDSS DR7 MPA-JHU catalog of \cite{Brinchmann2004}. The solid red line denotes the maximum location of starbursts by \cite{Kewley2001}, the dashed blue line is the star-forming boundary suggested by \cite{Kauffmann2003}, and the green line is the division between Seyfert galaxies and LINERS \citep{Kewley2006}. } 
         \label{fig:BPT}
   \end{figure*}

In Figure \ref{fig:BPT}, the three common BPT diagnostic plots are shown, and the flux ratios of Nucleus 1 and 2 are shown as a solid colored triangle and a circle, respectively. These are the flux ratios of the integrated spectra. We can only determine an upper limit for the flux of H$_{\beta}$ and S[II] in Nucleus 2, which is indicated as arrow. All other lines are significantly detected.

Nucleus 2 is solidly in the Seyfert region of all three panels in the BPT. This provides strong, independent evidence that Nucleus 2 is indeed hosting a SMBH and is also a low-luminosity AGN.  
The uncertainty in $H_{\beta}$ does not affect this conclusion, as we derive a \emph{\emph{lower}} limit; thus, the real [OIII]$\lambda5007$/H$_{\beta}$ line ratio might be even higher, which only puts it further into AGN territory in the diagnostic plots. With an X-ray luminosity of L=$2.8\times 10^{39}$\,erg\,s$^{-1}$ , Nucleus 2 is above the $10^{39}$\,erg\,s$^{-1}$ cut-off luminosity for low-mass X-ray binaries, which are extremely rare at higher luminosities. High-mass X-ray binaries can also cause such X-ray luminosities, but they only exist in stellar populations younger than 100Myr \citep{Fabrika2021}, of which we do not find any in our stellar population analysis.

It is also unlikely that the X-ray signal is caused by an accumulation of X-ray binaries that have sunken to the center due to mass segregation, as for this the relaxation timescale would have to be on the order of the age of the universe or less. We calculated the relaxation timescale in the center of Nucleus 2 based on Equation 8-71 from \cite{Binney1987}. Using a stellar density in the central 5pc of Nucleus 2 of $\rho=4\times10^3M_{\odot} pc^{-3}$, an average stellar mass of $0.7M_{\odot,}$ and a velocity dispersion of $\sigma=70km s^{-1}$, we derive a relaxation timescale of $t_{rel}=105\times10^{9}$yrs. With a relaxation time almost one order of magnitude larger than the age of the universe, it is unlikely that significant mass segregation has happened in the center of Nucleus 2.
Therefore, the combination of the optical emission lines, X-ray source, and our dynamical SMBH detection provide strong evidence that Nucleus 2 hosts a low-luminosity AGN. A typical value observed for the bolometric correction is $L_{\rm X}/L_{\rm bol}\sim16$ \citep{Ho2009}, and thus a rough estimate for Nucleus 2 provides $L_{\rm bol}=4.5\times10^{40}$\,ergs/s and therefore an Eddington accretion ratio of $L_{\rm bol}/L_{\rm edd}=3\times10^{-5}$. This puts Nucleus 2 in the typical range of observed low-luminosity Seyfert AGN.

Nucleus 1 is well within the AGN region of all three BPT diagnostic panels. Its integrated line ratio is right at the border between Seyfert galaxies and LINERs. However, due to the absence of an X-ray source in Nucleus 1 (see \citealt{Schweizer2018} for a discussion), we cannot be certain that the source of this ionizing radiation is the SMBH in its center. LINER emission can either be caused by a SMBH or the emission of hot stars in the post-AGB phase. These stars are common in very old stellar populations, as is the case for Nucleus 1. In the absence of an X-ray signal, Nucleus 1 cannot immediately be classified as an active SMBH. While it could be caused by the SMBH in its center, it is also possible that it is caused by the emission of hot post-AGB stars. Thus, only the SMBH in Nucleus 2 is actively accreting.

\section{Discussion}\label{sec:discussion}
In this work, we tested whether NGC\,7727 hosts a double SMBH system. This galaxy has a main nucleus at its photometric center, and a second object that appears to be a nucleus as well and is offset from the center by only 500\,pc. There was indirect evidence for a possible SMBH in both nuclei \citep{Schweizer2018}. We tested this hypothesis by using adaptive-optics-supported MUSE kinematical data centered on NG7727 to resolve the sphere of influence of both putative SMBHs and measured their masses. From our dynamical modeling, we find that both nuclei indeed host a SMBH. In this section, we discuss this result in the context of Nucleus 2 being a stripped former galaxy nucleus and the implications of this tight double SMBH system.

\begin{table}
\caption{ \label{tab:summary} Summary of the main results for Nuclei 1 and 2.}
\centering
\begin{tabular}{cccccc}
\hline\hline
Name  & Nucleus 1 & Nucleus 2 \\
\hline
R.A. & 23:39:53.796 &  23:39:53.679  \\
DEC.  & -12:17:34.04  &  -12:17:30.83  \\
$M_{\rm BH}$ [$M_{\odot}$] & $1.54^{+0.18}_{-0.15}\times10^{8}$   & $6.33^{+3.32}_{-1.40}\times10^{6}$   \\
$M_{\rm Bulge}$ [$M_{\odot}$] & $5.24 \times10^{10}$ & $2.10\times10^{8}$ \\
Integrated $\sigma$ [km\,s$^{-1}$] & 191.2 $\pm$ 1.5  &  66.3 $\pm$ 1.3 \\
Velocity [km\,s$^{-1}$] & 1794.9 $\pm$ 1.9  &   1839.2 $\pm$1.8 \\
\end{tabular}
\end{table}

\subsection{The nearest and closest dynamically confirmed dual SMBH system}

We detect the dynamical signature of SMBHs in both nuclei of NGC\,7727, confirming it as a close dual SMBH. In Nucleus 1, we found a BH with a mass of $M_{\rm BH}=1.54^{+0.18}_{-0.15} \times10^{8}M_{\odot}$ and in Nucleus 2 a SMBH with a mass of $M_{\rm BH}=6.33^{+3.32}_{-1.40}\times10^{6}M_{\odot}$. The two SMBHs have an apparent separation of $\sim500$\,pc. This makes them the closest known SMBH pair. All previous confirmed dual SMBHs have a separation of 1\,kpc or more \citep{Husemann2018, Kollatschny2020}. The SMBH in Nucleus 2 has 4.1\% of the mass compared to the central SMBH in Nucleus 1, and thus the SMBHs have a 1:24 mass ratio. 

In order to obtain an estimate for the inspiral time of Nucleus 2, we ran simulations of its orbit in the centre of NGC\,7727 under the influence of dynamical friction. We assumed a mass of Nucleus 2 of $2\times10^{8}M_{\odot}$, the current projected distance of 480\,pc between the nuclei and a relative velocity of 45\,km\,s$^{-1}$ between Nucleus 1 and 2 as starting conditions. We also assumed a range of z distances randomly distributed between 0 and 3\,kpc and random tangential velocities between 0 and 200\,km\,sec$^{-1}$. We find a typical merger time of 250\,Myr and that basically all cases merge within 1\,Gyr. Therefore, it is likely that this system merges soon and the SMBH-SMBH merger has a 1:24 mass ratio.

With its 27\,Mpc in distance, NGC\,7727 is the dual SMBH system closest to us. Before this work, the closest known dual SMBH was NGC\,6240 \citep{Komossa2003, Kollatschny2020}, which at 144\,Mpc is almost six times as far away from us as NGC\,7727. In NGC\,6240, similar MUSE data were used to study the system, but only the nuclear components could be spatially separated from each other. Resolving the sphere of influence dynamically at these distances is impossible with current IFU instrumentation. Typical catalogues of dual SMBHs/AGNs focus on systems that are even further away, usually at distances of $z>$0.2 ($\sim 800$\,Mpc). This illustrates how the proximity of NGC\,7727 at only 27\,Mpc allows us to study dual SMBHs in more detail than any other known system.

\subsection{Surviving nuclear star clusters are outliers to the black hole scaling relations}
\begin{figure*}
   \centering
   \includegraphics[width=\hsize]{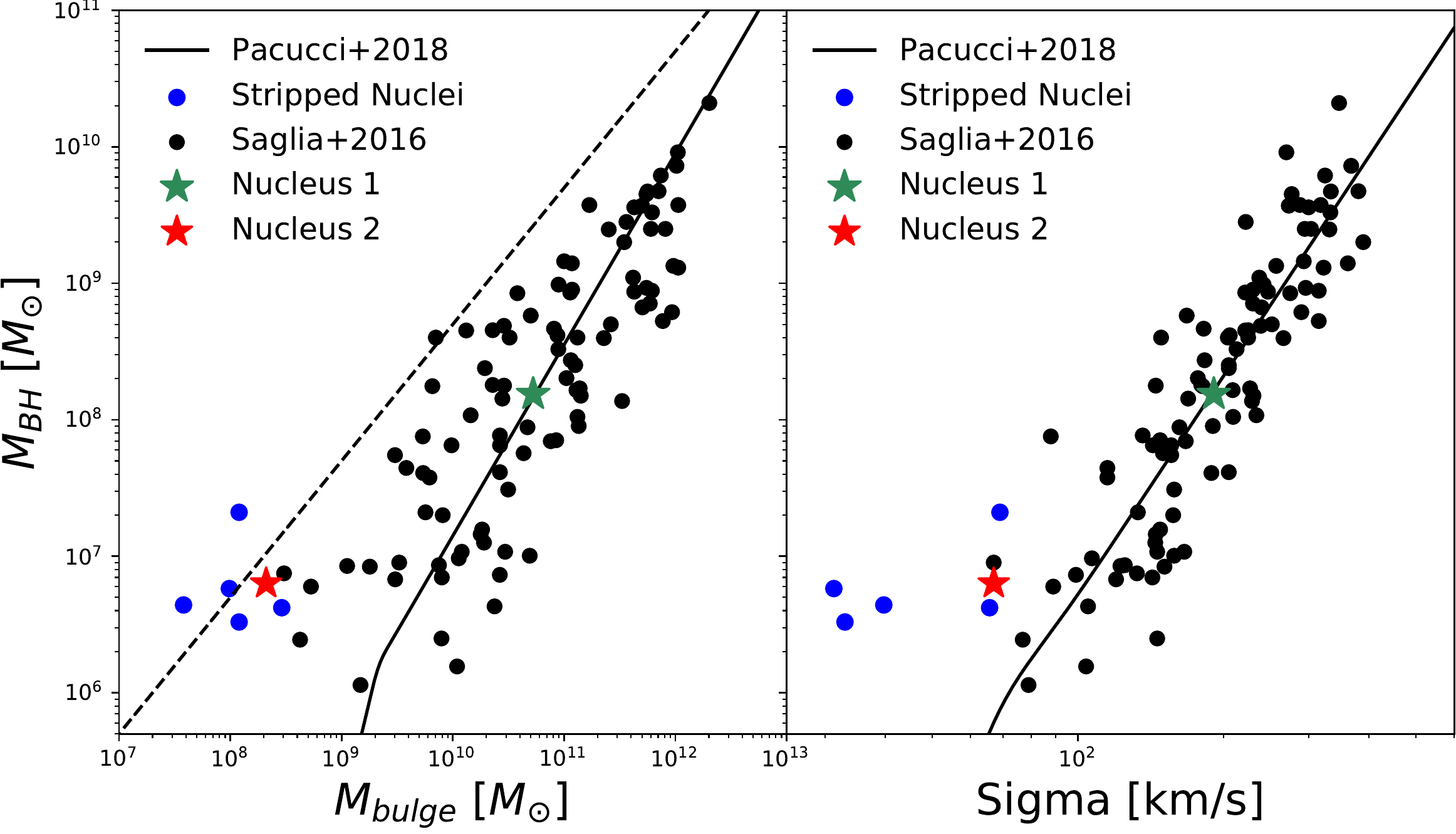}
      \caption{Black hole galaxy scaling relations. \textbf{Left Panel:} Bulge-$M_{\rm BH}$ relation with data points of \cite{Saglia2016} is shown in black, and the confirmed stripped nuclear star clusters with SMBHs are shown in blue \citep{Seth2014, Ahn2017, Ahn2018, Afanasiev2018}. Nucleus 1 is marked by a green star, where Nucleus 2 and its SMBH are marked by a red star symbol. The relation of \cite{Pacucci2018} is shown as a solid line in both panels, with a constant  SMBH mass fraction of 5\% shown by a dashed line in the left panel. \textbf{Right Panel:}Same datapoints as in the left panel but for the $M_{\rm BH}$ - sigma relation.  } 
         \label{fig:BH_scaling}
   \end{figure*}

Supermassive black holes and their stellar components obey two tight scaling relations, the $M_{\rm BH}-\sigma$ relation between the mass of the SMBH and the velocity dispersion of its bulge, and the one between the BH mass and bulge mass (see e.g., \citealt{Kormendy2013} for a detailed Review). In Fig. \ref{fig:BH_scaling}, we show the two BH scaling relations, with the data of the compilation of \citep{Saglia2016} in black and the five known stripped nuclear star clusters with SMBHs in blue \citep{Seth2014, Ahn2017, Ahn2018,Afanasiev2018}. The SMBHs in Nuclei 1 and 2 are shown in green and red, respectively. The black line is the relation from \cite{Pacucci2018}, and the dashed line indicates systems where the BH is exactly 5\% of the bulge mass.

The SMBH in Nucleus 1 falls right onto both the BH scaling relations, meaning that its SMBH is exactly as massive as predicted for its bulge mass of $M=5.24 \times10^{10}M_{\odot}$ and integrated velocity dispersion of $\sigma=191\rm \,km\,s^{-1}$. In contrast, Nucleus 2 is significantly offset to the left of the relation and falls exactly in the area of the other known surviving nuclear clusters with a SMBH (shown in blue).  All the surviving nuclear clusters with SMBHs have over-massive SMBHs compared to the M$_{\rm BH}$-Bulge relation, with BH mass fractions between 2 and 15\% of their total mass. Nucleus 2 hosts a similarly over-massive SMBH with 3.0\% of its total mass. This population of outliers with over-massive SMBHs, to the left of the scaling relations has been predicted theoretically in SMBH formation simulations when galaxies are tidally destroyed \citep{Volonteri2008, Volonteri2016, Barber2016}.

Other galaxies such as NGC\,3368 that fall close to the area of stripped nuclei in the scaling relations are large spiral galaxies with physical sizes of the order of kpc, who have an abnormally small (pseudo-)bulge and are thus relative outliers to the relation \citep{Nowak2010}. In contrast to these objects, stripped nuclei have typical physical sizes of 10-35\,pc, which are comparable to nuclear star clusters and not large galaxies.

 \subsection{Nucleus 2 as a former nuclear star cluster of a merged galaxy}The existence of a SMBH in Nucleus 2 is expected if it is the stripped nucleus of a merged galaxy. While there are only a handful of direct dynamical confirmations of SMBHs in surviving nuclei, a statistical analysis has recently shown that such relics of the merging process should be very common in the local Universe \citep{Voggel2019}.  The confirmation of stripped nuclei has been mainly based on the presence of a SMBH in their centers \citep{Seth2014, Ahn2017, Ahn2018, Afanasiev2018} that is so over-massive (2-15\% of the total mass) that it cannot be formed by any known stellar processes. The SMBH in Nucleus 2 is similar and makes up 3.0\% of the total mass of Nucleus 2, which puts it right in the typical mass range of other surviving star clusters \citep{Voggel2018}. However, the main difference compared to the other known surviving nuclei is that Nucleus 2 is in close proximity of the center of its host galaxy, whereas all other stripped nuclei are out in the halo, far from the center of their host galaxy at typical distances of $>6$\,kpc. In addition, Nucleus 2 is embedded in a blue tidal stream, and NGC\,7727 and its general disturbed morphology \citep{Schweizer2018} point to a recent merger. With the confirmation of the SMBH in Nucleus 2, the evidence is strong that Nucleus 2 has been caught in the process of being stripped from its host galaxy. Thus, it is a rare direct glimpse into the process of stripped nuclei formation as well as SMBH assembly through mergers.

Using the M$_{\rm BH}$-Bulge relation, we can estimate  the mass of the progenitor galaxy bulge to have been $M=5.2\times10^{9}M_{\odot}$. This indicates that the former host galaxy had a bulge that was one order of magnitude smaller than the bulge of NGC\,7727. However, this comparison does not take into account the mass of the disks in both galaxies.  To do this, we used the estimate of NGC\,7727 total stellar mass of $1.35\times10^{11}M_{\odot}$ \citep{Schweizer2018}. With the young stellar age of Nucleus 2 and the blue stellar stream, it is a reasonable assumption that the progenitor galaxy had a sizeable stellar disk that makes up at least twice the stellar mass of its bulge. This results in a total merger mass ratio of 1:5, indicating that it was a minor merger event that created the disturbed morphology of NGC\,7727. 

Additional evidence that Nucleus 2 is the accreted nucleus of a merged galaxy is its very blue color and young stellar population, which is at odds with it having formed within NGC\,7727, whose stellar population is uniformly red and old ($\sim$ 13\,Gyr). The several tidal streams and structures surrounding NGC\,7727 were found in \cite{Schweizer2018} and have the same blue color as Nucleus 2. Nucleus 2 itself is embedded in such a blue tidal stream. This strengthens the notion that the progenitor host of Nucleus 2 was a gas-rich young galaxy that merged with NGC\,7727 in the last 1-2\,Gyr.

\section{Conclusion}\label{sec:conclusion}
In this paper, we present adaptive-optics narrow-field MUSE IFU kinematic data on the two nuclei of NGC\,7727. Combining the kinematical data with HST photometry and Jeans anisotropic models, we can constrain whether there is a SMBH present in both nuclei.
Our main conclusions are summarized as follows.\begin{itemize}
\item[$\bullet$] We confirm a SMBH in the photometric center of NGC\,7727 with a mass of $M_{\rm BH}=1.54^{+0.18}_{-0.15}10^{8}M_{\odot}$. This is a first SMBH mass measurement for this galaxy. \\
\item[$\bullet$]  We confirm a second SMBH in Nucleus 2, which is offset from the center of the galaxy by 500\,pc in projected separation. This second BH has a mass of $M_{\rm BH}=6.33^{+3.32}_{-1.40}\times10^{6}M_{\odot}$. \\
\item[$\bullet$]  Our finding of a SMBH in Nucleus 2 that makes up 3.0\% of the underlying nucleus mass, confirms that it used to be the center of the galaxy that merged with NGC\,7727 and is a tidally stripped nuclear star cluster. Using the M$_{\rm BH}$-bulge relation, the bulge mass of its progenitor galaxy can be estimated at $M\sim5\times10^{9}M_{\odot}$. The progenitor galaxy of Nucleus 2 that merged with NGC\,7727 is likely a gas-rich disk galaxy, and assuming the disk had twice the mass of its bulge, it was a minor merger with a 1:5 mass ratio. \\
\item[$\bullet$]  The emission line profile in Nucleus 2 is similar to those of Seyfert AGNs. These optical emission lines in combination with the X-ray source in its center provide strong independent evidence that the SMBH in Nucleus 2 is a low-luminosity AGN. Nucleus 1 falls in the area of LINER galaxies, and due to the absence of an X-ray signal, the source of ionizing radiation is not immediately obvious. It could either be caused by the SMBH or by hot post-AGB stars. \\
\item[$\bullet$] This system constitutes the first dynamical detection of a dual SMBH system in which the stars in the sphere of influence have been resolved. In addition, this system has a separation of only 500\,pc, making it the only known SMBH pair with a sub-kpc separation. \\
\item[$\bullet$]  Its distance of only 27.4\,Mpc makes it the record holder for the nearest dual SMBH pair to us, with the next known dual SMBH being at a distance of 144\,Mpc.\\
\item[$\bullet$]  The orbit of Nucleus 2 can be used to estimate a merging timescale of $<$1\,Gyr. This double SMBH pair is thus likely to merge in the future and produce a gravitational wave event. The mass ratio of this future SMBH merger will be 1:24, allowing us to make the first precise prediction for the mass ratio of a SMBH merger. \\
\end{itemize}

It was already suggested that SMBHs in stripped nuclei are so common that they increase the SMBH density in the local Universe by up to 30\% \citep{Voggel2019}. The discovery of this offset SMBH is another confirmation that many SMBHs exist outside the centers of galaxies that have not been discovered yet. The search for surviving nuclear star clusters of accreted galaxies is currently the only manner to find these hidden SMBHs. The second nucleus of NGC\,7727 is an even more interesting object due to its small separation from the main SMBH in the center of NGC\,7727 and its advanced stage of merging. All five previous confirmed surviving nuclei with SMBHs were discovered further out in the halo of their host galaxy (at several kpc) and thus have much longer merging timescales.

This serendipitous discovery of a dual SMBH system studies offers a first view of a sub-kpc separation dual SMBH. This indicates that there are likely many more SMBHs as well as dual SMBH pairs in the local Universe that have been missed by surveys, as both are not actively accreting. Targeting bright surviving nuclear star clusters in merged galaxies can facilitate the discovery of dual SMBHs at smaller separations in the local Universe, even when they are not luminous AGNs. This will allow much more detailed studies of these systems that can then serve as blueprints of how to find them more broadly in the distant Universe.

The dynamical detection of SMBHs is observationally limited by the current capabilities of adaptive optics IFU instruments such as MUSE, to resolve the sphere of influence of a given SMBH. With the upcoming next generation of 30 meter telescopes and advanced instruments such as HARMONI on the ELT, the current distance limitations of this method will be pushed considerably. Currently, the typical MUSE AO PSF is $\sim0.1\arcsec$ with HARMONI; this will typically be 0.01-0.03$\arcsec$ \citep{Harmoni2008}, enabling us to resolve the sphere of influence of SMBHs for many more galaxies than currently possible.

\begin{acknowledgements}
We would like to thank Jay Strader, Sabine Thater and Nicolas Martin for helpful discussions concerning this paper.Work on this project by A.C.S. was supported by NSF grant AST-1813609.  Based on observations made with ESO Telescopes at the La Silla Paranal Observatory under programme ID 0103.B-0526(A). This research made use of Astropy,\footnote{http://www.astropy.org} a community-developed core Python package for Astronomy \citep{astropy:2013, astropy:2018}.
\end{acknowledgements}

\bibliographystyle{aa}
\bibliography{bib_NGC7727}

\newpage
\begin{appendix}
\section{Luminosity models}

 \begin{table}
\caption{ \label{tab:mge_nuc1} Luminosity model of Nucleus 1 from the multi-Gaussian expansion.}
\centering
\begin{tabular}{cccc}
\hline\hline
Luminosity & $\sigma$  & q & Position Angle \\
\hline
[$L_{\odot} pc^{-2}$] &  [arcsec]  &  & [Degrees] \\
\hline
20953.19 &  0.005 &     0.69  &    83.43 \\
31928.07   &     0.016 &      0.69  &   83.43 \\
37778.16 &     0.045 &       0.69  &    83.43 \\
32921.86 &      0.103  &      0.69  &    83.43 \\
19752.75  &       0.205  &       0.69  &   83.43 \\
7688.43  &      0.366  &       0.69 &     83.43 \\
1847.96 &       0.598  &       0.69  &    83.43 \\
257.51  &      0.923 &       0.69  &    83.43 \\
14.34  &      1.425  &      0.69  &     83.43 \\
\hline      
1308.94  &       0.347 &      0.69 &   -79.89 \\
2849.06  &       1.289  &   0.69  &  -79.89 \\
3448.48  &       3.040 &      0.69  &   -79.89 \\
2033.29  &       5.512 &     0.69   &  -79.89 \\
507.08  &       8.538 &      0.69  &   -79.89 \\
36.72  &      12.332 &      0.69 &     -79.89 \\
\hline
\end{tabular}
\end{table}

 \begin{table}
\caption{ \label{tab:mge_nuc2} Luminosity model of Nucleus 2 from the multi-Gaussian expansion. }
\centering
\begin{tabular}{cccc}
\hline\hline
Luminosity & $\sigma$  & q & Position Angle \\
\hline
[$L_{\odot} pc^{-2}$] &  [arcsec]  &  & [Degrees] \\
\hline
65426.46   &    0.002  &     0.62   &   70.51 \\
87892.47    &   0.007 &     0.62   &   70.51 \\
96648.39   &   0.019  &     0.62   &   70.51 \\
83412.83   &    0.045   &   0.62   &   70.51 \\
53812.23  &    0.094 &     0.62 &    70.51 \\
24566.65  &     0.178  &     0.62  &    70.51\\
7585.31   &    0.310  &     0.62   &   70.51 \\
1539.61    &   0.505   &    0.62   &   70.51 \\
194.19    &   0.787   &    0.62  &    70.51 \\
10.58    &   1.243   &    0.62   &   70.51 \\
\hline
\end{tabular}
\end{table}
\end{appendix}



\end{document}